\documentclass[fleqn]{2023SCGE}
\usepackage{url}
\usepackage{hyperref}
\usepackage{multirow}
\usepackage[normalem]{ulem}
\usepackage{makecell}

\newcommand{\farcs}{\mbox{$.\!\!^{\prime\prime}$}}

\def\ynk#1 {{\textcolor{cyan}{#1}}\ }

\begin{document}

\ensubject{subject}

%%%%%%%%%%%%%%%%%%%%%%%%%%%%%%%%%%%%%%%%%%%%%%%%%%%%%%%
%%% Authors do not modify the information below
%%% ????????????????
%%% ??????????, ????????????{}, ???????????????????
%Letter to the Editor??Article%??????
\ArticleType{Article}%??Article
\SpecialTopic{SPECIAL TOPIC: Astronomy}%???????
\Year{2024}
\Month{January}
\Vol{xxx}
\No{xx}
\DOI{xxx}
\ArtNo{000000}
\ReceiveDate{\today}
\AcceptDate{\today}
%\OnlineDate{January 1, 2016}
%%%%%%%%%%%%%%%%%%%%%%%%%%%%%%%%%%%%%%%%%%%%%%%%%%%%%%%

%%% title: ????
%%%   \title{title}{title for citation}
\title{CSST Preparations: Galaxy Completeness and Sérsic Profile Fitting across the Wide, Deep, and Extreme Fields}{CSST Preparations: Galaxy Completeness and Sérsic Profile Fitting across the Wide, Deep, and Extreme Fields}

\author[1]{Ziqi Ma}{}
\author[1]{Si-Yue Yu$^{\dag}$}{}
\author[1]{Taotao Fang$^{\ddag}$}{}
\author[2,3]{Jinyi Shangguan}{}
\author[4,5]{Zhao-Yu Li}{}
\author[2,3]{Luis C. Ho}{}

\footnote{$^{\dag}$ syu@xmu.edu.cn}
\footnote{$^{\ddag}$ fangt@xmu.edu.cn}

\AuthorMark{Ziqi Ma}
\AuthorCitation{Ziqi Ma,Si-Yue Yu, Taotao Fang, and et al.}

\address[{\rm1}]{Department of Astronomy, Xiamen University, Xiamen, Fujian 361005, P.R. China}
\address[{\rm2}]{Department of Astronomy, Peking University, Beijing 100871, China}
\address[{\rm3}]{Kavli Institute for Astronomy and Astrophysics, Peking University, Beijing 100871, China}
\address[{\rm4}]{Department of Astronomy, School of Physics and Astronomy, Shanghai Jiao Tong University, 800 Dongchuan Road, Shanghai, 200240, China}
\address[{\rm5}]{State Key Laboratory of Dark Matter Physics, School of Physics and Astronomy, Shanghai Jiao Tong University, Shanghai, 200240, China}

%\contributions{}%????????

%%% Abstract. ??
\abstract{The upcoming imaging survey of the Chinese Space-station Survey Telescope (CSST) will deliver high-resolution imaging of an unprecedented number of galaxies for galaxy studies. To understand CSST’s capability, and to support the preparation of early-science programs, we generate 470,526 mock CSST images for 22,406 simulated galaxies with $M_*>10^9\,M_\odot$, whose parameters are calibrated to match real HST observations spanning photometric redshift $0<z\lesssim7$, across seven CSST filters and three planned survey depths: wide, deep, and extreme. We then perform source detection and Sérsic fitting. For point sources, we found that the 95\% completeness magnitude in the {\it g} band reaches 26.3, 27.4, and 28.5\,mag for the wide, deep, and extreme fields, respectively.  For extended galaxies, their spatial extent dilutes the surface brightness, leading to brighter 95\% completeness magnitudes of 24.4, 25.9, and 27.1\,mag. The detection completeness remains above 95\% at $z\lesssim3$--4 in the extreme field, while the corresponding redshift limits are $z\approx1$ in the deep field and $z\approx0.5$ in the wide field. Using three fitting codes, {\tt GALFIT}, {\tt AstroPhot}, and {\tt SourceXtractor++}, we quantify measurement biases and uncertainties in galaxy magnitude ($m$), effective radius ($R_e$), effective surface brightness ($\mu_e$), Sérsic index ($n$), and axis ratio ($q$).  On average, for fainter galaxies, the reduced signal-to-noise ratio leads to systematic overestimates in $m$, $R_e$, and $\mu_e$, and underestimates in $n$ and $q$. These biases, as well as the associated scatter, become progressively smaller in deeper fields. Overall, our results provide quantitative constraints on sample selection and the robustness of morphological measurements in CSST early-science and legacy surveys.}

%%% Keywords. ?????
\keywords{Galaxies, CSST galaxy survey, galaxy morphology, galaxy photometry}
%%%%%%%%% https://s3-us-west-2.amazonaws.com/clarivate-scholarone-prod-us-west-2-s1m-public/wwwRoot/prod3/societyimages/scpma/pacs.pdf
\PACS{95.55.-n, 95.55.Fw, 95.75.-z, 98.52.-b}
%%%%%%%%% 95.55.-n  Astronomical and space-research instrumentation
%%%%%%%%% 95.55.Fw  Space-based ultraviolet, optical, and infrared telescopes
%%%%%%%%% 95.75.-z  Observation and data reduction techniques; computer modeling and simulation
%%%%%%%%% 98.52.-b  Normal galaxies; extragalactic objects and systems (by type)
%%%%%%%%% 
%%%%%%%%%
%%%%%%%%% 
%%%%%%%%% 
%%%%%%%%% 
%%%%%%%%% 
%%%%%%%%% 
\maketitle

%\tableofcontents%?????

%%%%%%%%%%%%%%%%%%%%%%%%%%%%%%%%%%%%%%%%%%%%%%%%%%%%%%%
%%% The main text. ???????
%???????????????????\cref{fig1}
%\twocolumn\onecolumn
%%%%%%%%%%%%%%%%%%%%%%%%%%%%%%%%%%%%%%%%%%%%%%%%%%%%%%%
\begin{multicols}{2}

\section{Introduction}
The Chinese Space-station Survey Telescope (CSST) is a space-based survey facility designed to conduct wide-field, high-resolution imaging and spectroscopic observations from the near-ultraviolet (NUV) to the {\it y} band. CSST is an off-axis three-mirror anastigmat telescope with a 2-m primary mirror and a $f/14$ Cooke-type optical design, optimized to deliver stable, near-diffraction-limited image quality over a large field of view \cite{Zhan2021}. CSST is designed to operate in the same orbit as the China Manned Space Station and can dock with the station for on-orbit maintenance and instrument servicing. This unique operational mode enables long-term, stable survey operations while preserving instrument performance throughout the mission lifetime. 

CSST is equipped with five scientific instruments: the Multi-band Imaging and Slitless Spectroscopy Survey Camera (hereafter the Survey Camera, SC), the Multi-Channel Imager, the Integral Field Spectrograph, the Cool Planet Imaging Coronagraph, and the THz Spectrometer \cite{Yan2025IFU,Yan2025MCI,Wei2025Mock,Zhu2025CPIC,Tan2025HSTDM,Zhao2025CPIC,Wei2025Overview,Xian2025,Ban2025,Xie2025,Feng2025,Zhang2025}. Among these, the SC is designed to carry out the primary mission of CSST, and the majority of the observing time will be allocated to this instrument. The SC performs wide-field multi-band imaging and slitless spectroscopic surveys, with a science focal plane comprising 2.6 billion pixels and a field of view of 1.1 square degrees.

During its nominal ten-year mission, the CSST imaging survey is planned to include multiple levels. The wide-field survey will cover 17,500 square degrees in seven imaging bands and three low-resolution spectroscopic bands, with an integrated exposure time of $t_{\rm exp}=300$\,s. In addition, deep-field observations with $t_{\rm exp}=2000$\,s will be performed over 400 square degrees in the same set of bands. An extreme-field survey, reaching $t_{\rm exp}=15{,}000$\,s, will further cover 10 square degrees, providing ultra-deep imaging for selected regions. 
The primary goals of CSST include precision cosmology, through measurements of the Universe's expansion history, dark energy properties, and the distribution of dark matter; and studies of galaxy and active galactic nuclei (AGN) evolution, by probing galaxy morphology, mass assembly, and the co-evolution of supermassive black holes and their host galaxies. Additional objectives encompass detailed studies of the Milky Way and nearby galaxies, stellar structure and evolution, detection and characterization of exoplanets, small bodies in the Solar System, precise astrometry, and monitoring of transients and variable sources.

Despite the planned capabilities of CSST, a systematic, quantitative evaluation of galaxy detection completeness across its multiple survey tiers has not yet been validated. Such an assessment is essential for enabling robust statistical studies of galaxy populations. 
The different imaging strategies of CSST will probe galaxy populations to different depths, leading to distinct detection limits. Incomplete detection at fixed stellar mass or luminosity can introduce significant selection biases and compromise studies of cosmic evolution \cite{PracticalStatistics2012}, particularly at high redshift. Establishing the completeness limits for each survey tier is therefore essential for defining robust galaxy samples and for identifying the redshift ranges over which key populations can be reliably studied. 

In addition, a robust measurement of galaxy structural parameters is essential for the scientific goals of CSST. Modeling fitting based on Sérsic function is widely used to measure galaxy fluxes and morphologies in large imaging surveys \cite{Shen2003, Simard2011, Shuntov2025}. By explicitly accounting for PSF convolution, Sérsic fitting enables consistent photometric and structural measurements across images with different resolutions and observing conditions, which is particularly important for CSST’s multi-band imaging.
Sérsic fitting provides key morphological parameters such as effective radius, Sérsic index, and surface brightness, which are fundamental for studies of galaxy structure and evolution \cite{vanderWel2014,Kartaltepe2023}. However, the robustness of these measurements depends on observation depth and image resolution. For marginally resolved or noisy galaxy images, fitting results can be biased and may not accurately reflect intrinsic galaxy properties \cite{Yu2023}.
Measurement biases are commonly quantified using mock observations constructed either from idealized Sérsic models \cite{Meert2013, Rowe2015Galsim, Davari2016, Davari2017, Euclid2023XXVI, Lzy2025, Wei2025Mock} or from high-quality images of nearby galaxies, re-observed under the conditions of a given survey \cite{Giavalisco1996,Conselice2003,Yu2018,Yu2023,Liang2024}. Mock galaxies based on Sérsic models are particularly well suited for this purpose, as they allow controlled exploration of a wide range of intrinsic parameters and galaxy populations.

Given the unprecedented combination of area, resolution, and depth of the CSST imaging survey, a pre-launch assessment of the robustness of Sérsic fitting is timely and necessary \cite{Zhan2021, Wei2025Overview}. 
In this work, we use extensive suites of mock galaxy images, constructed with Sérsic models, to quantify both galaxy detection completeness and systematic fitting errors across the three planned CSST survey depths (wide, deep, and extreme), employing multi-code validation to ensure robust results under the expected observing conditions.
This study provides essential guidance for early CSST galaxy science and for the statistical interpretation of future CSST imaging data.

This paper is organized as follows. In Section~\ref{sec:sampledata}, we describe the construction of the mock galaxy catalog and images. Section~\ref{sec:method} outlines the measurement methods. Section~\ref{sec:results} presents the main results and discussion, and Section~\ref{sec:conclusion} summarizes our conclusions. Throughout this paper, we adopt AB magnitudes, a Ref. \cite{Chabrier2003} initial mass function, and assume a flat $\Lambda$CDM cosmology with $(\Omega_{\rm m}, \Omega_{\Lambda}, h) = (0.27, 0.73, 0.7)$.

\section{Mock Image Creation}  
\label{sec:sampledata}
For this study, we generate simulated galaxy images to construct mock datasets for the CSST/SC\footnote{The CSST filter set has been recently updated, with the {\it y} band removed and two new filters (WV and WI) introduced. The image simulations presented here remain sufficient to demonstrate the methodology; future applications should adopt the updated configuration.}, spanning seven photometric bands (NUV, {\it u}, {\it g}, {\it r}, {\it i}, {\it z}, and {\it y}) across the three planned survey depths (wide, deep, and extreme field). For each band, the system throughput ($T_{\rm filter}$), taking into account the detector quantum efficiency, optical transmission, and filter transmission (shown in Figure~\ref{fig:filter}), as well as the point-spread functions (PSFs; whose radial profiles are shown in Figure~\ref{fig:psf}), are adopted from the CSST image simulation software\footnote{\url{https://csst-tb.bao.ac.cn/code/csst-sims/csst_msc_sim}} developed by Ref. \cite{Wei2025Mock}. The exposure-time configurations defining the three survey depths are summarized in Table~\ref{tab:exptime}. The effective exposure time is doubled in the NUV and {\it y} bands due to the detector design.

\subsection{Generation of Galaxy Catalog} \label{sec:2.1}
We generate mock galaxy catalogs using the Empirical Galaxy Generator (EGG)\cite{Schreiber2017}. EGG is an empirically driven framework calibrated on the observed properties of galaxies detected in the CANDELS fields \cite{Grogin2011, Koekemoer2011, Guo2013, Galametz2013, Nayyeri2017}. The photometry adopted in EGG combines space- and ground-based imaging and typically covers wavelengths from the {\it U} band to the 8\,$\mu$m, with additional mid- and far-infrared data from {\it Spitzer} and {\it Herschel} when available. The redshifts used in EGG are photometric redshifts. In EGG, redshifts ($z$), stellar masses ($M_\ast$), and rest-frame $U-V$ and $V-J$ colors of galaxies are drawn from observed properties of galaxies in CANDELS. Other physical properties, including star formation rates, dust attenuation, and morphology, are assigned based on empirical relations. The bulge-to-total ratio ($B/T$) was assigned using the observed relations between the average $B/T$ and stellar mass for both star-forming and quenched galaxies in the CANDELS fields across a range of redshifts, as reported by Ref. \cite{Lang2014}.  We note that, in the EGG catalog, elliptical galaxies have $B/T$ values close to, but not exactly, unity. For each simulated galaxy, based on its position on the {\it U}{\it V}{\it J} diagram, a panchromatic spectral energy distribution (SED) is constructed, and synthetic photometry is obtained by integrating the redshifted SED over broad-band filters, with the CSST filter set adopted as input. The resulting mock catalogs reproduce observed galaxy number counts and morphological properties, including the cosmic evolution of galaxy sizes. The average imaging depth of GOODS-S and GOODS-N fields \cite{Grogin2011, Koekemoer2011} is deeper than that expected for the CSST extreme field. Combined with the broad wavelength coverage, this makes EGG well suited to providing realistic mock catalogs that closely match the anticipated CSST observations.

EGG generates mock CSST galaxy images by assigning effective radius to galaxies drawn from the Jiutian simulation \cite{Han2025}, adopting effective radius at $z=0$ measured by Ref. \cite{Zhang2019RAA} and assuming an evolution model. In contrast, the EGG catalog is directly calibrated to observed HST galaxy size–mass relations across redshift \cite{Schreiber2017}. To ensure consistency with the observed size distributions of CANDELS galaxies, we therefore adopt the EGG galaxy catalog in this work.

Using EGG, we generate a mock catalog of 22,406 galaxies with $M_* > 10^9\,M_\odot$ over an area of $300~\mathrm{arcmin}^2$, centered at ${\rm RA}=150^\circ1$ and ${\rm Dec}=2^\circ2$. For each mock galaxy, the EGG catalog provides separate bulge and disk components. The disk component is modeled with an exponential radial profile (Sérsic index $n_{\rm disk}=1$), while the bulge component follows a de Vaucouleurs profile ($n_{\rm bulge}=4$). For each component and each band, the catalog reports the total flux as well as the structural parameters, including the effective (half-light) radius, axis ratio, and position angle. The effective radius is defined as the radius enclosing half of the total flux of the corresponding component, the axis ratio characterizes its projected ellipticity, and position angle specifies the orientation of the major axis on the sky.

Since single-Sérsic fitting is widely used in survey-based morphological analyses \cite{Shen2003, Simard2011, Shuntov2025}, image simulations are typically constructed using single-Sérsic models to enable a quantitative assessment of measurement biases and uncertainties \cite{Meert2013, Hiemer2014, Rowe2015Galsim, Euclid2023XXVI, Wei2025Mock}. In contrast, using double-Sérsic models makes the definition of parameters such as the single-Sérsic index less well defined. Therefore, following the strategy in Ref. \cite{Euclid2023XXVI} and \cite{Euclid2023Merlin}, we convert the two-component model into a single-Sérsic model.

\begin{figure*}
	\centering
	\includegraphics[width=2\columnwidth]{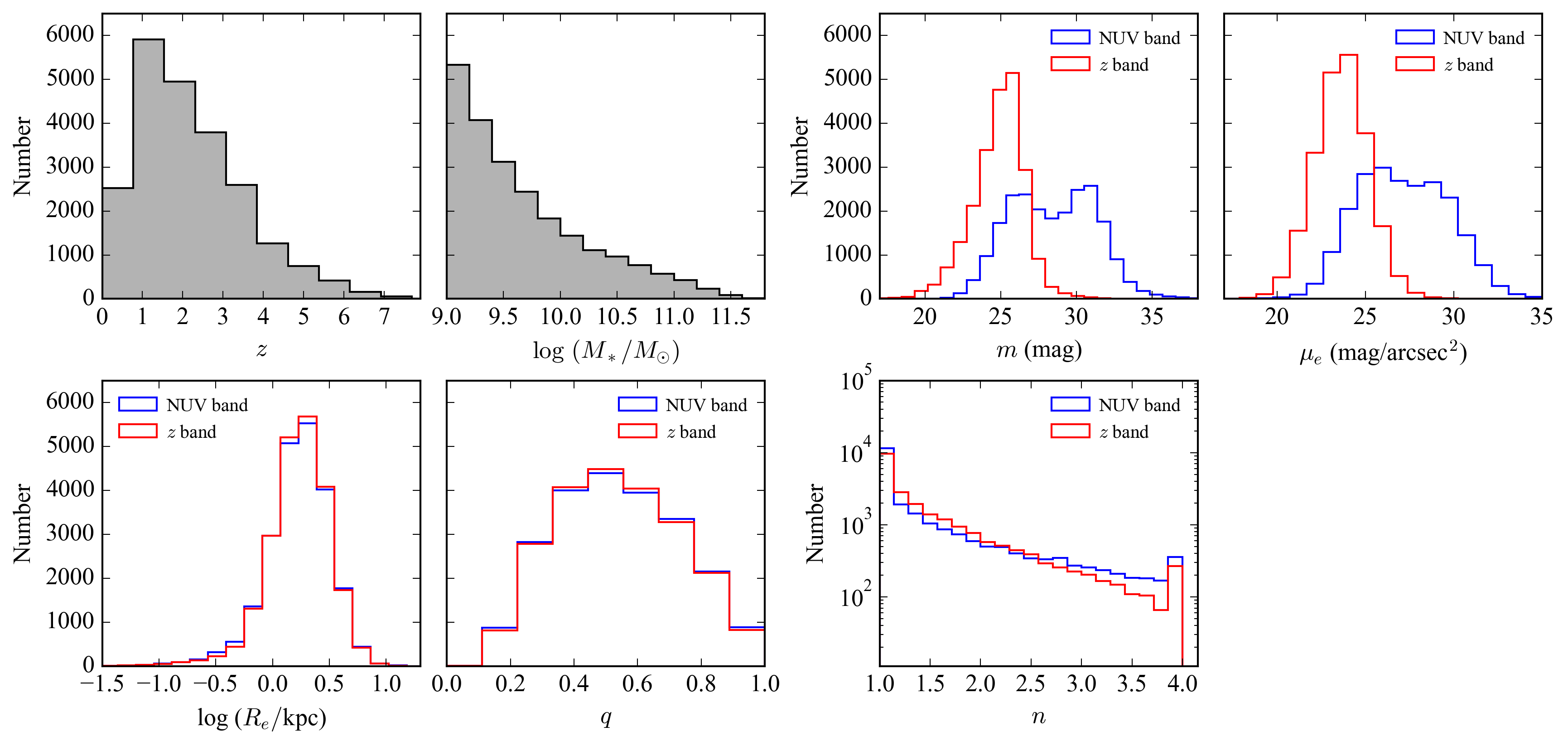}
	\caption{Distributions of single-Sérsic model parameters in the simulated sample. Top row shows redshift ($z$), stellar mass ($M_*$), apparent magnitude ($m$), and effective surface brightness ($\mu_e$); bottom row shows effective radius ($R_e$), axis ratio ($q$), and Sérsic index ($n$).
	}
	\label{fig:properties}
\end{figure*}

For each galaxy, we first compute the total flux in each band. The remaining structural parameters of the single-component model are then determined based on the bulge-to-total flux ratio ($B/T$), following the prescriptions described in Ref. \cite{Schreiber2017} and \cite{Euclid2023XXVI}. The Sérsic index $n$ of the single-component model is defined as
\begin{equation}
	n = n_{\rm disk}\,(1-B/T) + n_{\rm bulge}\,B/T,
\end{equation}
which naturally yields $n=1$ for purely disk-dominated systems ($B/T=0$) and $n=4$ for purely bulge-dominated systems ($B/T=1$). The upper limit of $n$ is therefore 4. The axis ratio $q$ and position angle (PA) are computed in the same manner:
\begin{equation}
	q = q_{\rm disk}\,(1-B/T) + q_{\rm bulge}\,B/T
\end{equation}
and
\begin{equation}
	{\rm PA} = {\rm PA}_{\rm disk}\,(1-B/T) + {\rm PA}_{\rm bulge}\,B/T.
\end{equation}
The intrinsic effective radius ($R_e$) is then determined numerically through isophotal analysis with {\tt Python} package {\tt Photutils} \cite{larry_bradley_2025}, using elliptical isophotes defined by the derived $q$ and PA. Based on the resulting structural parameters, the surface brightness at the effective radius, $I_e$, is computed analytically. The final single-Sérsic surface brightness profile \cite{Sersic1968} is thus given by
\begin{equation} \label{eq:sersic}
	I(R) = I_e \, \exp \Big\{ -b_n \Big[ \Big( \frac{R}{R_e} \Big)^{1/n} - 1 \Big] \Big\},
\end{equation}
\noindent
where the $b_n$ is a function of $n$ determined by the incomplete gamma function.  The distributions of the single-Sérsic model parameters in our simulated sample, including redshift ($z$), stellar mass ($M_*$), apparent magnitude ($m$), effective radius ($R_e$), effective surface brightness ($\mu_e=m+2.5\log(2\pi R_e^2)$), axis ratio ($q$), and Sérsic index ($n$), are summarized in Figure~\ref{fig:properties}.
As demonstrated in Ref. \cite{Euclid2023XXVI}, the resulting single-Sérsic parameters reproduce the observed property distributions in the COSMOS field, although the models exhibit a slightly higher fraction of galaxies with relatively low Sérsic indexes.

Since no NUV photometry is included in the construction of the EGG catalog, the NUV fluxes are inferred from spectral energy distributions constrained by optical colors. While very faint NUV emission is expected for quiescent systems, the resulting NUV fluxes in EGG can reach levels of $\sim\,35$\,mag that are well below the detection limits of current facilities, as illustrated in Figure~\ref{fig:properties}.
Due to the fact that the NUV fluxes in EGG are inferred in the absence of direct NUV constraints, they may not fully reflect the true observational properties and could be systematically underestimated, particularly if unexpected sources are present. Direct NUV measurements from the CSST large-field survey in the near future will therefore be essential for a robust characterization of the NUV emission.

\subsection{CSST Observational Conditions} \label{sec:2.2}

\begin{figure}[H]
	\centering
	\includegraphics[width=1\columnwidth]{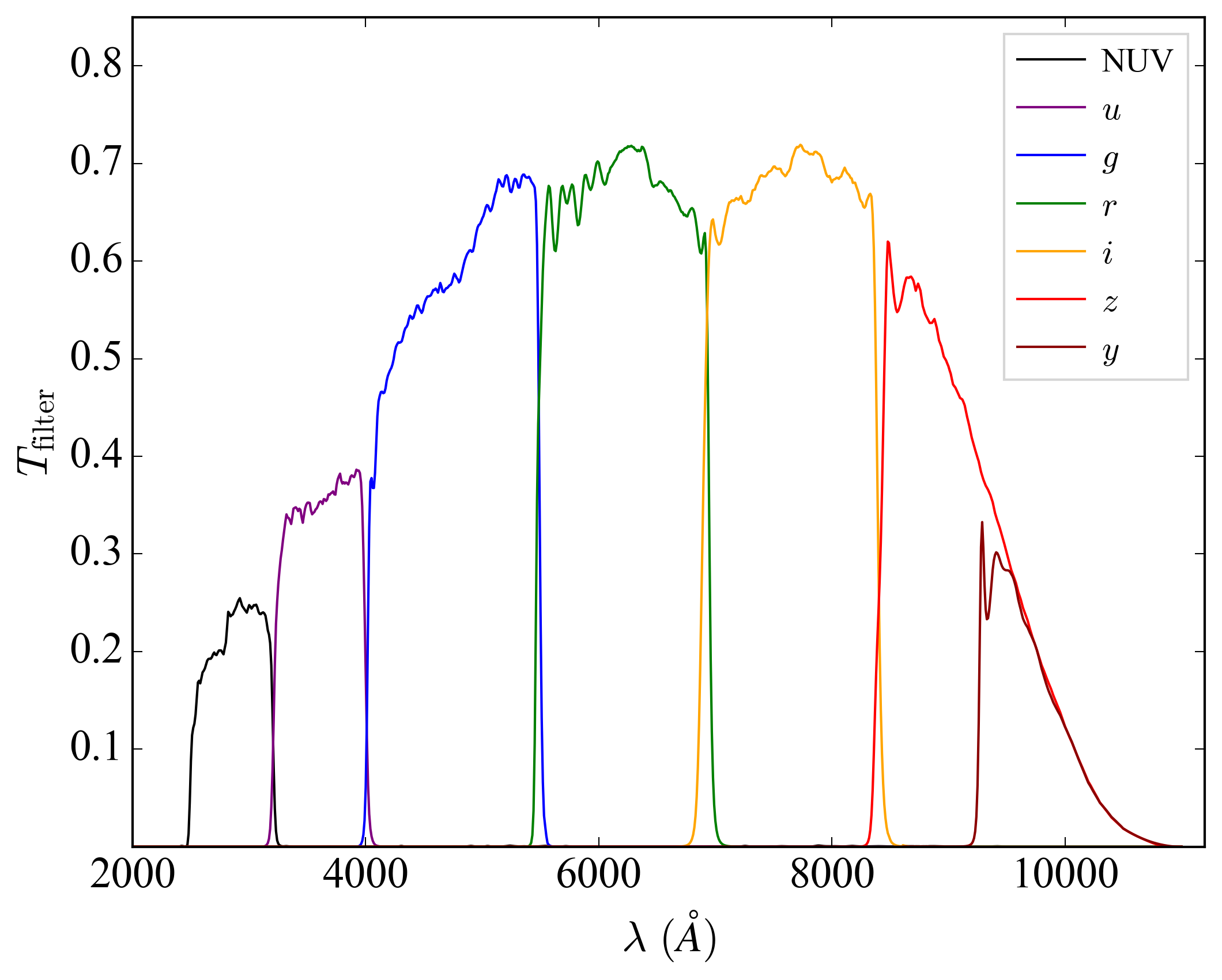}
	\caption{System throughput ($T_{\rm filter}$) of the CSST survey camera filters from NUV to the {\it y} band.}
	\label{fig:filter}
\end{figure}

\begin{figure*}
	\centering
	\includegraphics[width=2\columnwidth]{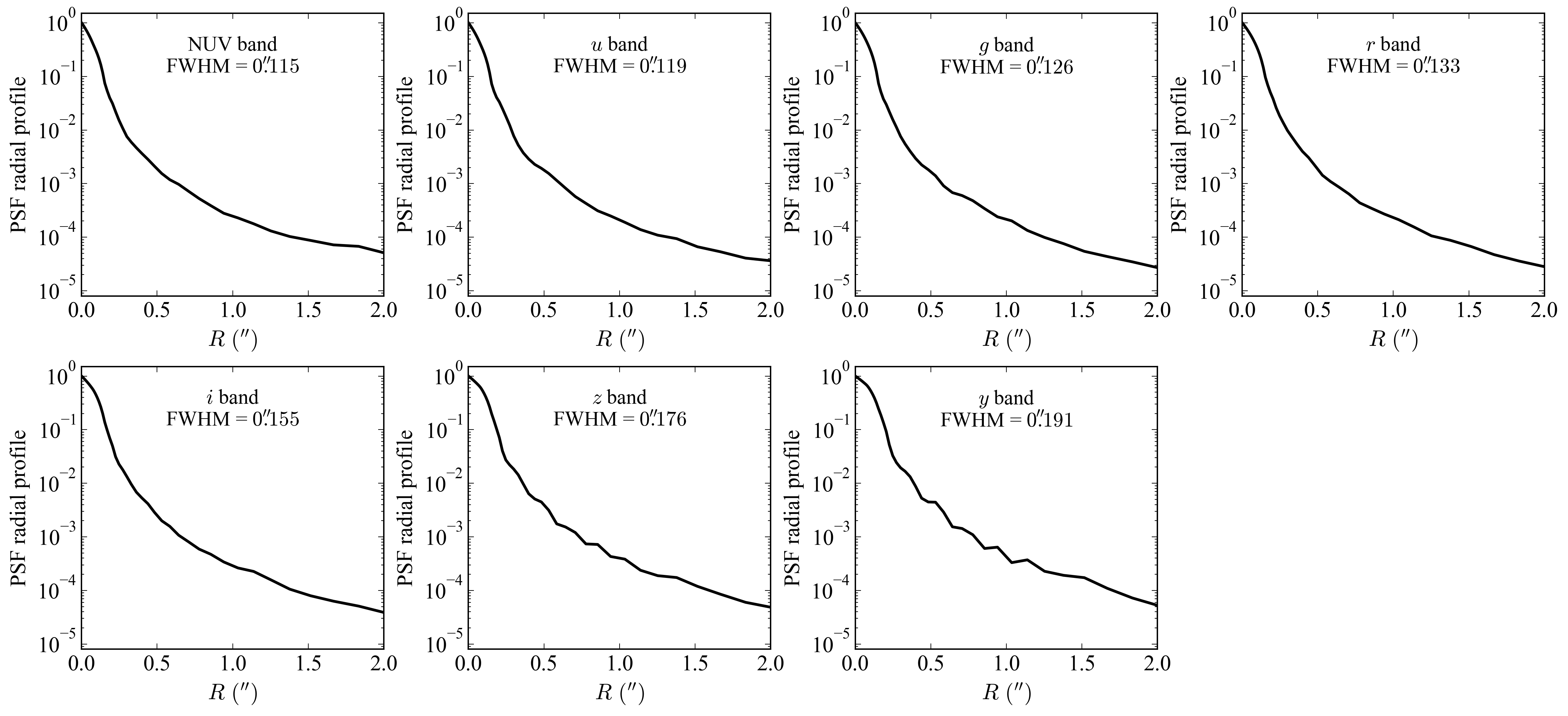}
	\caption{Radial profiles of the PSF in bands from NUV to {\it y}.
		The corresponding full width at half maximum (FWHM) are listed in each panel.
	} 
	\label{fig:psf}
\end{figure*}

The CSST/SC is a multiband imaging survey instrument, with a detector pixel scale of $\theta_\mathrm{pix}=0\farcs074$ per pixel and seven standard photometric channels (NUV, {\it u}, {\it g}, {\it r}, {\it i}, {\it z}, and {\it y}; see Figure~\ref{fig:filter}) that cover an effective wavelength range of 255--1000\,nm \cite{Wei2025Overview}. For each band, the radial profile sampled from the PSF is shown in Figure~\ref{fig:psf}. 
The PSF full width at half maximum (FWHM) is computed by fitting a Moffat function to mitigate the effects of pixel integration.
The resulted FWHM is listed in each panel, providing a quantitative measure of the image resolution in each band. 
The FWHM ranges from $0\farcs115$ at NUV band, to $0\farcs133$ at {\it r} band, and to $0\farcs191$ at {\it y} band. The three survey depths, referred to as the wide, deep, and extreme surveys, whose exposure configurations are summarized in Table~\ref{tab:exptime}.

\begin{table*}[htbp]
	\centering
	\caption{Exposure time configuration for different survey strategies in each band.}
	\label{tab:exptime}
	\begin{tabular}{cccccccc}
		\hline
		& NUV & $u$ & $g$ & $r$ & $i$ & $z$ & $y$ \\
		\hline
		Exposure time (Wide) 
		& $4\times150$ s 
		& $2\times150$ s 
		& $2\times150$ s 
		& $2\times150$ s 
		& $2\times150$ s 
		& $2\times150$ s 
		& $4\times150$ s \\
		Exposure time (Deep) 
		& $16\times250$ s 
		& $8\times250$ s 
		& $8\times250$ s 
		& $8\times250$ s 
		& $8\times250$ s 
		& $8\times250$ s 
		& $16\times250$ s \\
		Exposure time (Extreme) 
		& $120\times250$ s 
		& $60\times250$ s 
		& $60\times250$ s 
		& $60\times250$ s 
		& $60\times250$ s 
		& $60\times250$ s 
		& $120\times250$ s \\
		\hline
	\end{tabular}
\end{table*}

To generate simulated CSST/SC images, we first compute the photometric zero point for each filter in the AB magnitude system.
In this system, the AB magnitude is defined as
\begin{equation} \label{eq:5}
	m_\mathrm{AB} = -2.5 \log\left(   \frac{F_\nu}{3631\,{\rm Jy}}  \right),
\end{equation}
where $F_\nu$ is the flux density of the sources in units of
$\mathrm{erg\,cm^{-2}\,s^{-1}\,Hz^{-1}}$.  By converting to the unit of $\mathrm{erg\,cm^{-2}\,s^{-1}\,\AA^{-1}}$, we have the flux density:
\begin{equation} \label{eq:6}
	F_\lambda  = \frac{c}{\lambda^2} F_\nu ,
\end{equation}
where $c$ is the speed of light. The energy of a photon at $\lambda$ is
\begin{equation}
	E_{\lambda,\,\mathrm{photon}} = \frac{h c}{\lambda} ,
\end{equation}
where $h$ is the Planck's constant. The photon rate at $\lambda$ per unit collecting area is
\begin{equation}
	N_{\lambda,\,\mathrm{photon} }
	= \frac{F_\lambda}{  E_{\lambda,\,\mathrm{photon}} } .
\end{equation}
For a given $T_{\rm filter}(\lambda)$ and by assuming $F_\nu$ does not significantly change within the filter coverage, 
the observed photon rate is
\begin{equation} \label{eq:9}
	N_\mathrm{photon}
	= \frac{\pi}{h} \left(\frac{D}{2}\right)^2 \,F_\nu\,\int \frac{T_{\rm filter}(\lambda)}{ \lambda} d\lambda ,
\end{equation}
where $D=2\,$m, the diameter of the CSST primary mirror.
Therefore, the observed count rate yields
\begin{equation} \label{eq:10}
	N_{\rm count} = \frac{N_{\rm electron}}{G} = \frac{N_{\rm photon}}{G},
\end{equation}
\noindent
where $G$ is the detector gain and is set to 1 electron per count. The photometric zero point is defined as the magnitude of a source that produces
one count per second:
\begin{equation} \label{eq:11}
	{\rm zero~point} = -2.5 \log\left(   \frac{ F_\nu \big|_{N_{\rm count}=1}
	}{3631\,{\rm Jy}}  \right).
\end{equation}
\noindent
Combining Eq.~(\ref{eq:9}), (\ref{eq:10}), and (\ref{eq:11}), we calculate the photometry zero points for each band and  listed them in Table~\ref{tab:zps}.

We next consider the sky background, which includes contributions from natural sky emission and stray light, primarily arising from zodiacal light, earthshine, and geocoronal emission. The sky background is assumed to be flat and uniform. We adopt wavelength-dependent typically average background flux densities per unit solid angle ($F_{\lambda,\,\Omega}^\mathrm{bkg}$) derived from HST/WFC3 observations\footnote{\url{https://hst-docs.stsci.edu/wfc3ihb/chapter-9-wfc3-exposure-time-calculation/9-7-sky-background}}, as CSST operates in a low-Earth orbit similar to that of HST, making the HST-based sky background a valid approximation.

The background photon count rate per unit collecting area and per arcsec$^2$ is
\begin{equation}
	N_{\lambda,\,\Omega,\,\mathrm{photon}}^\mathrm{bkg}
	= \frac{F_{\lambda,\,\Omega}^\mathrm{bkg}}{E_{\lambda,\,\mathrm{photon} } }.
\end{equation}
The background photon rate per pixel is
\begin{equation}
	N_{\lambda,\,\mathrm{pix},\,\mathrm{photon}}^\mathrm{bkg}
	= \pi\, \left(\frac{D}{2}\right)^2 \,N_{\lambda,\,\Omega,\,\mathrm{photon}}^\mathrm{bkg}\, \theta_\mathrm{pix}^2.
\end{equation}
Finally, the background count rate per pixel is
\begin{equation}
	N^\mathrm{bkg}_{\rm pix,\,count} = \int N_{\lambda,\,\mathrm{pix},\,\mathrm{photon}}^\mathrm{bkg} \, T_{\rm filter}(\lambda) \, d\lambda .
\end{equation}

In addition to the sky background, we account for detector readout noise and dark current, adopting values from the CSST performance documentation\footnote{\url{https://nao.cas.cn/csst/research/performance/202508/t20250812_7903494.html}}. The readout noise, expressed in units of e$^-$/pix, quantifies the standard deviation of the electron counts added to each pixel during readout, while the dark current, in units of e$^-$/pix/s, represents the rate of thermally generated electrons in the absence of incident light. The photometric zero points, sky-background electron count rates, readout noise, and dark current for each band are summarized in Table~\ref{tab:zps}. 
We note that the adopted dark current value ($0.02$ e$^{-}$/pix/s) corresponds to the conservative upper limit assumed in the instrument design. For the fabricated CCDs, the measured dark current is typically lower, of order $0.001$ e$^{-}$/pix/s. 
Adopting $0.001$ e$^{-}$/pix/s reduces the total background noise by less than 3\%, for instance in the r-band deep field. Therefore, the exact value of the dark current does not significantly affect our results.

\begin{table}[H]
	\centering
	\setlength{\tabcolsep}{4pt}
	\caption{Photometric zero points, sky background, and detector parameters adopted in the simulations.}
	\label{tab:zps}
	\begin{tabular}{ccccc}
		\hline
		Filter & \shortstack{Photometric \\ zero point} & \shortstack{Background \\ count rate} & \shortstack{Readout \\ noise} & \shortstack{Dark \\ current} \\
		& (mag) & (e$^{-}$/s) & (e$^{-}$/pix) & (e$^{-}$/pix/s) \\
		\hline
		NUV & 24.91 & 0.0036 & 5 & 0.02 \\
		$u$  & 25.30 & 0.0186 & 5 & 0.02 \\
		$g$  & 26.24 & 0.1461 & 5 & 0.02 \\
		$r$  & 26.11 & 0.2124 & 5 & 0.02 \\
		$i$  & 25.91 & 0.2140 & 5 & 0.02 \\
		$z$  & 25.26 & 0.1269 & 5 & 0.02 \\
		$y$  & 23.90 & 0.0369 & 5 & 0.02 \\
		\hline
	\end{tabular}
\end{table}

\subsection{Simulation of CSST Galaxy Images} \label{sec:2.3}

For each galaxy in each band, the CSST galaxy image is simulated following the steps described below. We first construct a galaxy model described by a Sérsic function (Equation~\ref{eq:sersic}), with structural parameters determined in Section~\ref{sec:2.1}, using the Python package \texttt{AstroPhot} \cite{Stone2023}.  The model flux is then integrated over the total exposure time, corresponding to the addition of multiple individual exposures listed in Table~\ref{tab:exptime}, and the physical fluxes are converted into detector electron counts using the appropriate photometric zero point.  Instrumental effects are incorporated by convolving the model image with the PSF. By considering the total exposure, we then add Poisson noise associated with the combined signal from the galaxy, sky background, and dark current, and Gaussian readout noise to represent detector readout uncertainties. The readout noise, $5\,{\rm e^-}/{\rm pix}$, occurs in every individual exposure, and hence its integration grows linearly with the square-root of the number of exposures. For instance, in the deep field with 8 exposures, the standard deviation of the combined readout noise is $5\sqrt{8}\,{\rm e^-}/{\rm pix}=14\,{\rm e^-}/{\rm pix}$. The final simulated image is obtained by dividing the accumulated electron counts by the total exposure time, yielding images in units of e$^{-}/$s. In total, we have 470,526 mock CSST images for 22,406 mock galaxies. 

We note that here we consider idealized noise with a flat sky. In real observations, however, the sky background is not flat, and its accurate measurement is important for determining the effective observation depth, especially for low surface brightness galaxies. In addition, the drizzling process used in the co-addition of multiple individual exposures introduces pixel-level convolution, resulting in correlated noise. This changes the noise patterns and slightly degrades the CSST detection capability. Addressing these effects will require image simulations that incorporate a realistic data reduction process in future work.

A sigma map, representing the $1\sigma$ uncertainty of the pixel values in the simulated galaxy image, is required for Sérsic model fitting. To mimic the procedure adopted in real image analysis, we estimate the contribution from the galaxy flux by first replacing negative pixel values in the simulated image with zero, and then converting the image into a sigma map by assuming the Gaussian approximation to Poisson noise associated with the galaxy signal. The uncertainty due to other noise sources is estimated as the standard deviation measured in sky regions of the simulated image. These components are then combined in quadrature to obtain the final sigma map.

\subsection{Impact of PSF Undersampling} \label{sec:2.4}

As shown in Figure~\ref{fig:psf}, the PSFs among the seven bands have FWHM ranging from 0.115 arcsec to 0.191 arcsec, corresponding to 1.55--2.58 pixels at a pixel scale of 0.074 arcsec/pixel. According to the Nyquist criterion, PSFs narrower than 2 pixels are considered undersampled. To evaluate the impact of using an undersampled PSF on the recovery of galaxy structural parameters, we select the NUV band, which has the smallest FWHM, for a dedicated test.

We first generate two projected PSFs based on the best-fit Moffat function:
one projected at the native pixel scale with $\text{FWHM}=1.65$ pixels (hereafter the native PSF) and the other oversampled by a factor of two with $\text{FWHM}=3.3$ pixels (hereafter the oversampled PSF). Both PSFs adopt the same analytic Moffat parameters and differ only in their sampling. 
We generate a set of simulated galaxies modeled with Sérsic profiles. The effective radius $R_e$ is sampled at ten uniformly spaced values ranging from 0.5 to 3.0 times the native PSF FWHM. For each radius, Sérsic indices $n = 1, 2, 3$, and $4$ are considered. The magnitude is fixed at $m = 22$, the axis ratio at $q = 0.8$, and the position angle at $\theta = 0^\circ$.
To minimize the influence of noise on the comparison, noise corresponding to the extreme survey depth is added to the convolved images, which are subsequently fitted.

\begin{figure*}
	\centering
	\includegraphics[width=2\columnwidth]{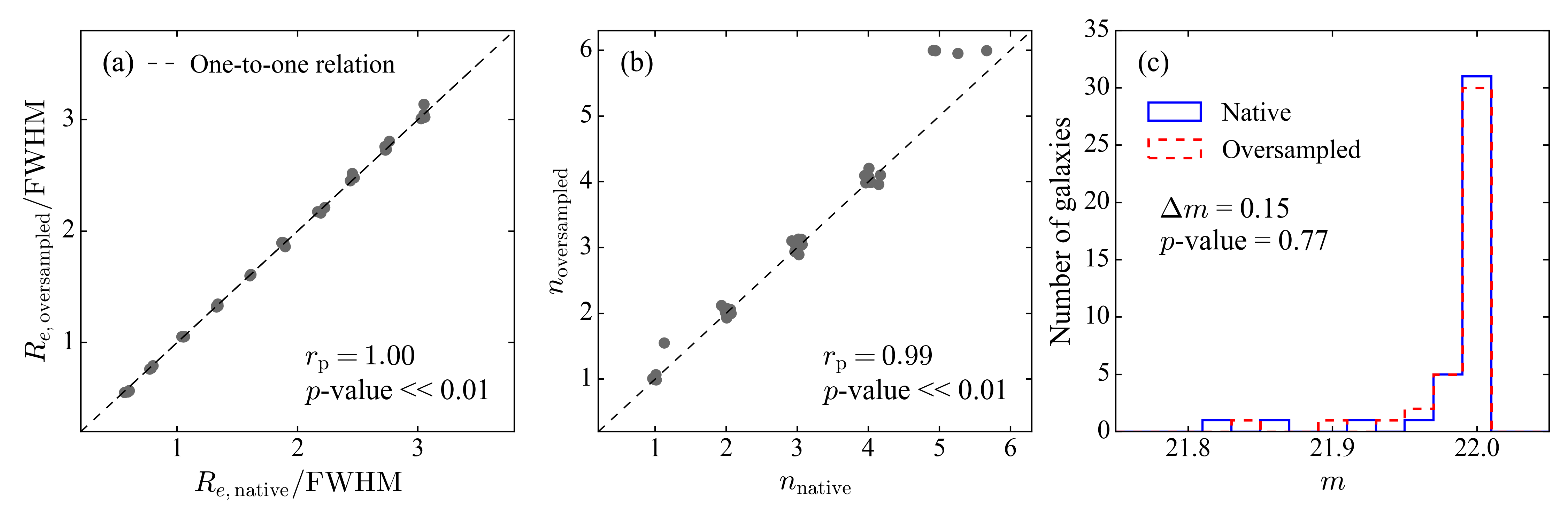}
	\caption{Comparison of the fitted structural parameters for simulated galaxy images convolved with the native PSF (original NUV-band PSF sampling, $\sim$1.55 pixels) and the oversampled PSF ($\sim$3.1 pixels) in the NUV band. 
	(a) Effective radius $R_e$ recovered from the two cases, expressed in units of the PSF FWHM. The dashed line indicates the one-to-one relation. The Pearson correlation coefficient $r$ is listed in the lower-right corner. 
	(b) Comparison of the recovered Sérsic index $n$ (same as panel~a). 
	(c) Comparison of the recovered magnitude distributions, with the Kolmogorov--Smirnov (KS) test statistic indicated in the lower-right corner.
	}
	\label{fig:psf_test}
\end{figure*}

The best-fit parameters from the two cases are compared in Figure~\ref{fig:psf_test}.
Figure~\ref{fig:psf_test}(a) shows the comparison of the recovered effective radii $R_e$, normalized by the FWHM, obtained using the native and oversampled PSFs. The Pearson correlation coefficient is $r_{\rm p} = 1.00$ with a $p$-value $\ll 0.01$, indicating a very strong and statistically significant linear correlation. The data points lie close to the one-to-one relation line, indicating that the best-fit $R_e$ values are nearly identical in the two cases. This result indicates that PSF undersampling has a negligible impact on the measurement of the galaxy effective radius in our simulations.
Figure~\ref{fig:psf_test}(b) shows the comparison for the Sérsic index \(n\).  The Pearson correlation coefficient is \(r_{\rm p} = 0.99\), and the corresponding \(p\)-value is \(\ll 0.01\), demonstrating a very strong correlation between the two cases. The results obtained using the native and oversampled PSFs are again highly consistent, with most points distributed close to the one-to-one relation. Nevertheless, when $n$ is high, measurements based on the native PSF tend to be slightly lower than those obtained with the oversampled PSF.
Figure~\ref{fig:psf_test}(c) compares the distributions of the recovered magnitudes. In this case we use the Kolmogorov--Smirnov (KS) test to determine whether the two samples are drawn from the same parent distribution. The KS statistic is $\Delta m = 0.15$ with a $p$-value of 0.77, indicating that the magnitude distributions recovered using the native and oversampled PSFs are statistically consistent, with no significant differences in our simulations.

Overall, for the simulated galaxies considered here, the structural parameters recovered after convolving the galaxies with the native and oversampled PSFs in the NUV band are very similar. Given that NUV has the smallest FWHM among the seven bands, this suggests that PSF undersampling is unlikely to introduce significant biases in structural parameter measurements for the other bands. 

\section{Image Photometry analysis} \label{sec:method}

\begin{figure}[H]
	\centering
	\includegraphics[width=1\columnwidth]{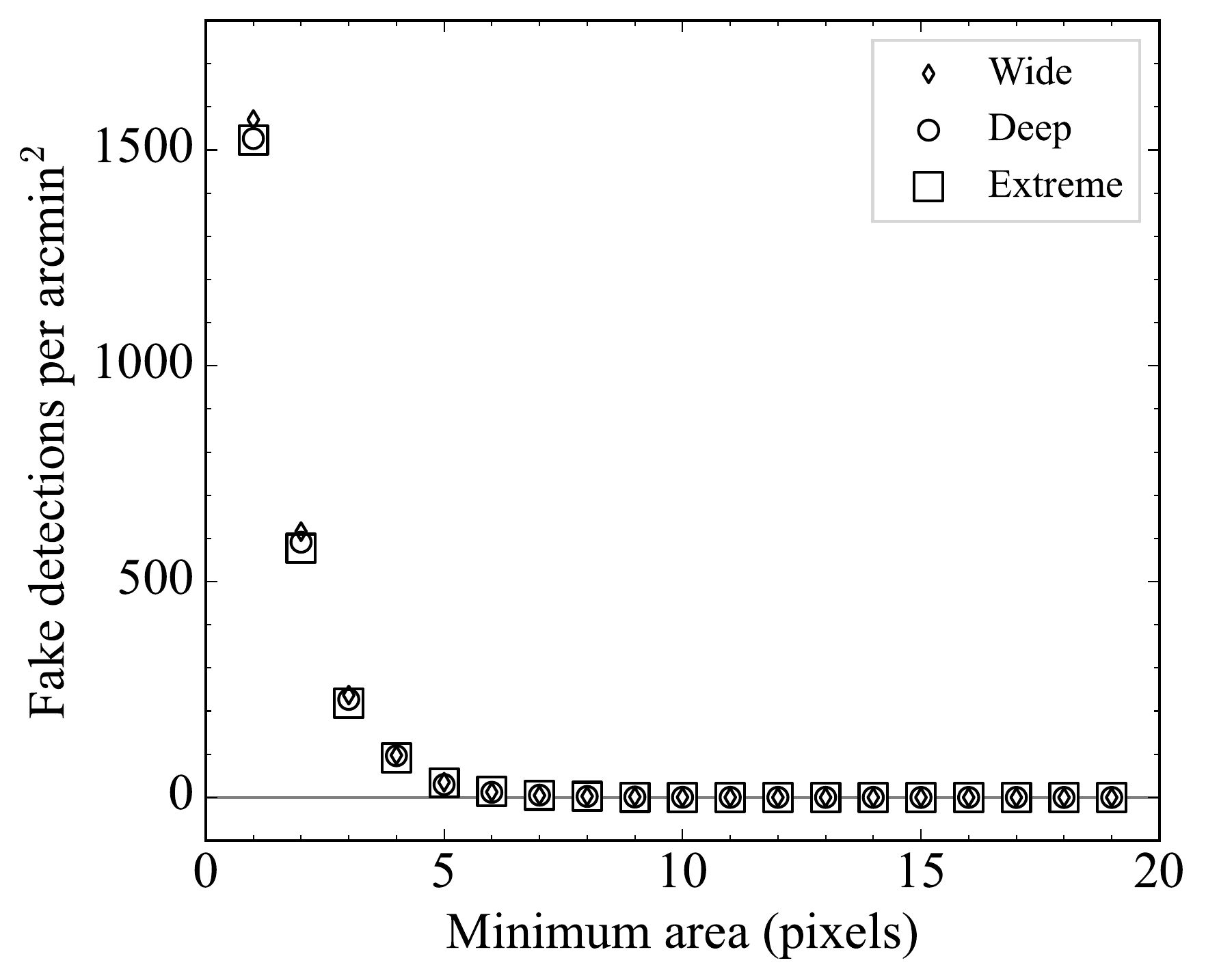}
	\caption{
		Fake detections per arcmin$^2$ (810 by 810 pixels) as a function of the minimum detection area. 
		Results are derived from a noise-only simulation for different survey strategies: the wide, deep, and extreme fields.
	}
	\label{fig:sep}
\end{figure}

In imaging analyses, source detection is typically performed prior to structural modeling, in order to identify astrophysical sources, extract postage-stamp images, and subsequently fit parametric models to each detected object. 
To detect galaxies as thoroughly as possible, we set the detection threshold in {\tt SEP} to $1\sigma$ above the background noise. Such a low threshold, however, can also lead to the identification of noisy pixels as spurious sources.
Another key parameter in source detection is the minimum detection area, which defines the smallest number of contiguous pixels required for a group of pixels to be classified as a source. Pixel groups smaller than this threshold are interpreted as noise fluctuations and are discarded, thereby suppressing spurious detections arising from isolated high-noise pixels. In this work, source detection is carried out using \texttt{SEP} \cite{Bertin1996, Barbary2016}.
To determine an appropriate minimum detection area, we generate noise-only images and evaluate detection performance for minimum areas ranging from 1 to 19 pixels. Figure~\ref{fig:sep} presents the surface density of false detections (per arcmin$^{2}$, measured in images of $810\times810$ pixels) as a function of the minimum detection area for the wide, deep, and extreme survey tiers. As expected, the false-detection rate decreases monotonically with increasing minimum area. We find that adopting a minimum area of 7 pixels reduces the number of spurious detections to a sufficiently low level, with negligible differences among the three survey depths. We therefore adopt a minimum detection area of 7 pixels throughout our analysis.
In practice, the minimum detection area is often chosen to approximate the area enclosed within one PSF FWHM; for example, this corresponds to 10 pixels in the CSST {\it g} band. We have verified that adopting a slightly larger minimum area (e.g., 10 pixels) produces nearly identical detection results, indicating that our conclusions are insensitive to the exact choice of this parameter.

We note several caveats regarding our source-detection procedure. In our simulations, each galaxy is modeled and analyzed in an independent image for each band, rather than being placed into a common sky frame according to their simulated celestial coordinates. As a result, source detection is performed without accounting for source blending. In real observations, galaxies are embedded in a shared sky background, where faint sources located near bright, extended galaxies may be partially suppressed. In addition, the background in real data is not perfectly uniform but can exhibit spatial variations; uncertainties in background estimation can further affect the detectability of faint sources, particularly those with low surface brightness. These effects are not included in the present analysis, and our results should therefore be interpreted as representing an optimistic, effectively upper-limit estimate of the detection and measurement performance achievable with CSST imaging under idealized conditions.

Additional complications arise in crowded environments, such as galaxy groups and clusters, where source detection becomes more challenging and typically requires careful tuning of detection and deblending parameters (e.g., \cite{Guo2013}). In such cases, the completeness may be reduced due to suppression of faint sources by nearby bright objects. Conversely, substructures within galaxies, especially spiral arms and clumpy star-forming regions, may be misidentified as independent sources. A full assessment of these effects, as well as a dedicated treatment of low surface brightness galaxies, is beyond the scope of this work and will be explored in future studies.

Galaxies that are not detected in the source-extraction stage are excluded from subsequent analysis. For the detected sample, we perform Sérsic model fitting using three independent software packages: \texttt{GALFIT}, \texttt{AstroPhot}, and \texttt{SourceXtractor++}. Employing multiple fitting codes allows us to assess potential systematic differences arising from the distinct optimization strategies, model implementations, and algorithmic assumptions adopted by each package.

{\tt GALFIT} \cite{Peng2002,Peng2010} is the most widely adopted two-dimensional parametric fitting package designed to simultaneously model one or more objects within an image. It fits analytic surface-brightness profiles to galaxy images and, in its simplest form, models galaxies with ellipsoidal components. More generally, it allows for flexible modeling of complex galaxy morphologies by combining multiple parametric components within a single fit.

{\tt AstroPhot} \cite{Stone2023} is a Python-based modular and object-oriented astronomical image modeling package. Its modular design allows individual components of the code to be modified or replaced without affecting other parts, making it flexible for a wide range of photometry applications. As an object-oriented package, AstroPhot enables intuitive manipulation of image and model objects. A key feature of AstroPhot is its class hierarchy of model objects, which allows users to customize models with minimal code modification. This structure facilitates building complex models for galaxy images by extending or combining existing components while maintaining full control over model parameters.

{\tt SourceXtractor++} \cite{Bertin2020} is a ground-up rewrite of SExtractor \cite{Bertin1996} written in C++, featuring a two-stage workflow of source detection and measurement. The latter is controlled by a flexible Python-based configuration and model-fitting framework that supports single- or multi-band analysis, allows multiple structural components to be fitted simultaneously, and adopts empirically calibrated parameter priors derived from comparisons between output distributions and the corresponding true catalog distributions.

For all fitting codes, we adopt a consistent fitting setup in order to enable a fair comparison between different methods. Each galaxy is fitted with a single Sérsic model, using the corresponding PSF and the sigma map constructed from the simulated image. The input images are simulated CSST galaxy image. The sigma maps are provided to the fitting codes as pixel-wise uncertainty estimates and are used to define the weighting of the likelihood or $\chi^2$ function. In all cases, the fitted parameters include the total flux or equivalently the flux at $R_e$, $R_e$, $n$, $q$, and PA.

\section{Results and Discussion}  \label{sec:results}

Galaxy detection completeness provides essential information on the faintest sources that can be reliably identified at different imaging depths, and is therefore critical for the preparation of CSST science prior to launch. In addition to detection completeness, the robustness of flux and structural-parameter measurements quantifies the accuracy and uncertainty with which galaxy properties can be recovered from CSST imaging data. Such assessments provide a practical basis for characterizing and correcting measurement biases induced by image degradation in future CSST survey analysis. In this section, we first examine the galaxy detection completeness as a function of redshift $z$ and apparent magnitude $m$ in Section~\ref{subsec:detection}, and then, in Section~\ref{sec:morph}, quantify the robustness of flux and morphological measurements by comparing best-fit parameters derived from the mock images with the true input values of the simulated galaxies. 

\subsection{Detection Completeness Analysis}  \label{subsec:detection}

\begin{figure*}
	\centering
	\includegraphics[width=2\columnwidth]{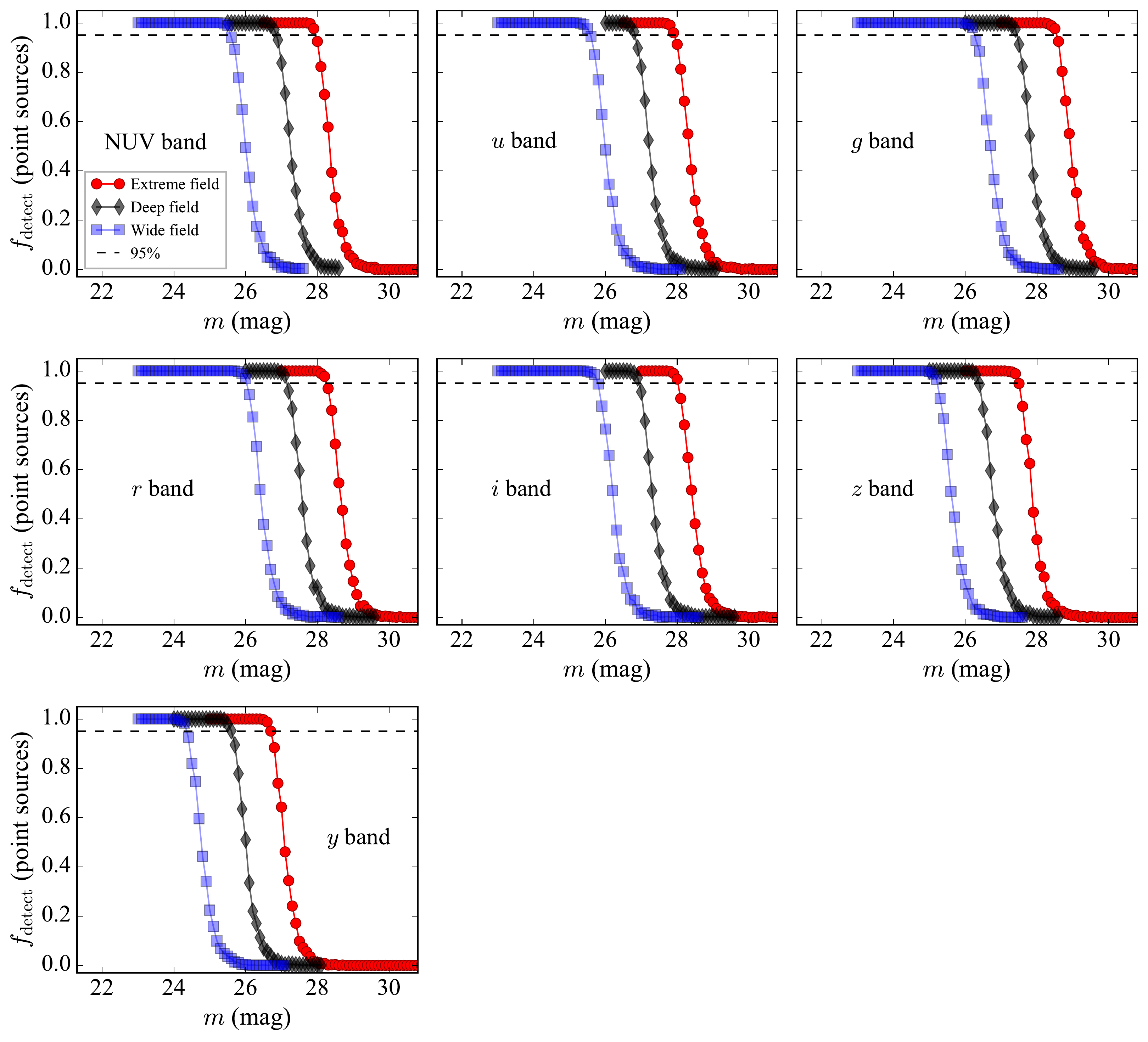}
	\caption{
		Point-source detection completeness ($f_{\mathrm{detect}}$) as a function of apparent magnitude ($m$) in different bands.  
		The results for the wide, deep, and extreme fields are denoted by blue squares, black diamonds, and red points, respectively. The horizon black dashed line indicates a detection completeness of 95\%.
	}
	\label{fig:fp_m}
\end{figure*}

We quantify the detection performance of the simulated CSST/SC observations using the detection completeness, $f_{\mathrm{detect}}$.  All objects in our simulations are generated and detected in isolation, and source confusion or blending effects are therefore not included.  We first evaluate detection limits for point sources. We generate a suite of simulated point sources spanning a range of apparent magnitudes and perform source detection as described in Section~\ref{sec:method}. Figure~\ref{fig:fp_m} shows that $f_{\mathrm{detect}}$ declines rapidly beyond a critical apparent magnitude toward fainter sources. We define the completeness magnitude as the magnitude at which $f_{\mathrm{detect}}=0.95$, and derive this quantity for each filter and survey depth. The resulting completeness magnitudes, corresponding to the 95\% point-source detection limits, are listed in Table~\ref{tab:detection_limit}. As expected, deeper survey strategies achieve systematically fainter detection limits in all filters. The typical difference in completeness magnitude is $\sim$\,1.1\,mag between the wide and deep fields, and a similar offset between the deep and extreme fields. In particular, in the {\it g} band, the completeness magnitude improves from 26.3\,mag in the wide field to 27.4\,mag in the deep field, and further to 28.5\,mag in the extreme field. Our derived limiting magnitudes based on detection completeness for the wide and deep survey are fully consistent with the 5\,$\sigma$ limiting magnitudes reported by Ref. \cite{Zhan2021}, with differences of $\lesssim$\,0.1\,mag (the extreme field was not considered in \cite{Zhan2021}). Our analysis provides a practical verification of the limiting magnitude calculations for CSST imaging.

\begin{figure*}
	\centering
	\includegraphics[width=2\columnwidth]{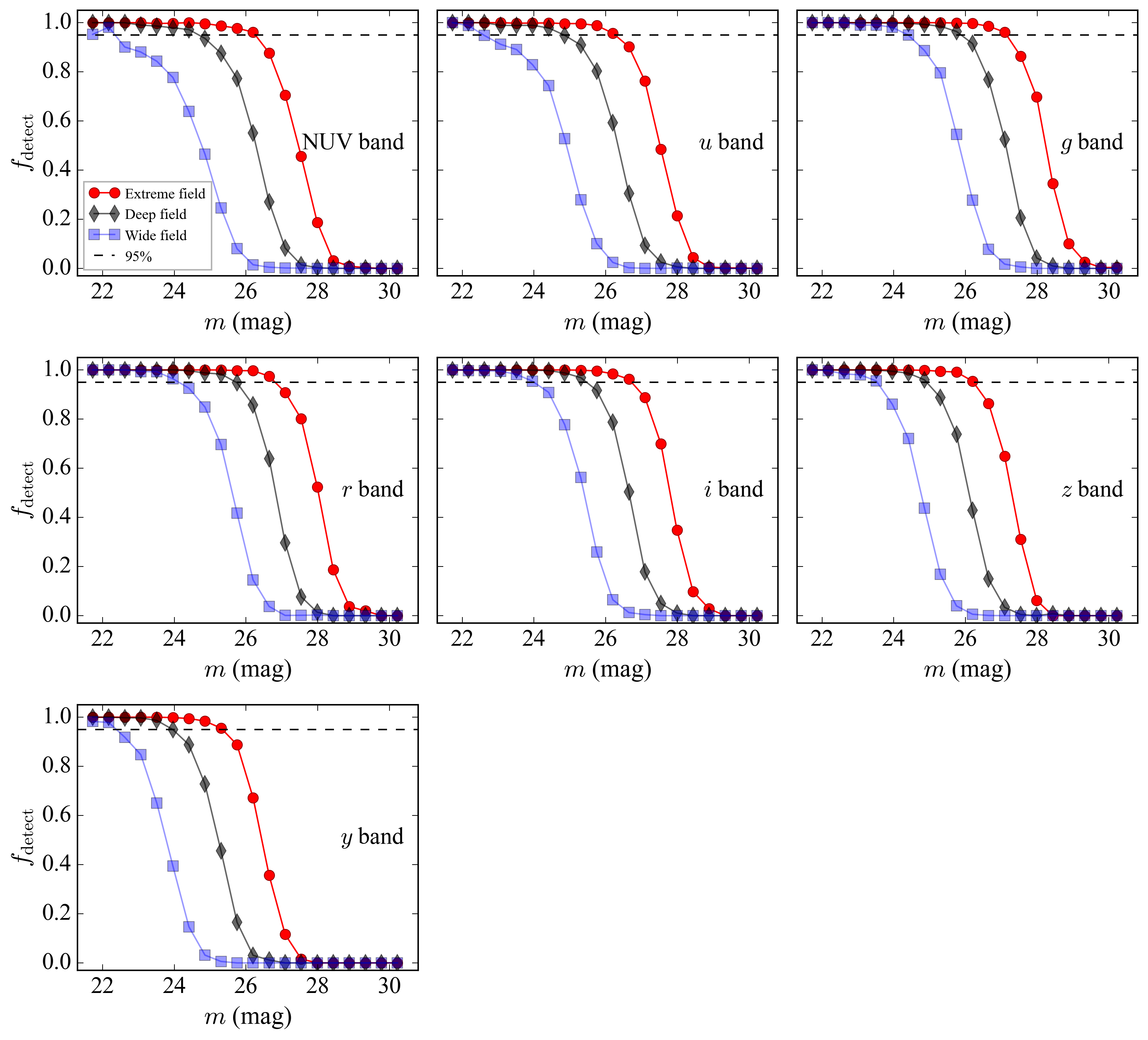}
	\caption{
		Detection completeness ($f_{\mathrm{detect}}$) for galaxies with $M_*>10^9\,M_\odot$ as a function of apparent magnitude ($m$) in different bands. 
		The panel layout and plotting conventions are the same as in Fig.~\ref{fig:fp_m}.
	}
	\label{fig:f_m}
\end{figure*}

We present the galaxy detection completeness for the simulated CSST/SC images in Figure~\ref{fig:f_m}. As in the case of point sources, the detection completeness $f_{\mathrm{detect}}$ remains close to unity for bright galaxies and decreases toward fainter apparent magnitudes beyond a characteristic threshold. However, the decline in completeness for galaxies is more gradual than that observed for point sources. The fundamental reason is the combined effect of PSF convolution and the fact that the radial profile of the galaxies, especially disk galaxies, is less centrally concentrated in the PSF.  Together with the PSF effect, galaxy flux is distributed over multiple pixels, resulting in lower per-pixel signal-to-noise ratio(S/N) at a fixed total apparent magnitude. Consequently, galaxy images are more susceptible to noise fluctuations and are more likely to fall below the detection threshold than point sources of the same apparent magnitude. The completeness magnitude, corresponding to the 95\% detection limits for galaxies with $M_*>10^9\,M_\odot$, at each band over each survey depth, are listed in Table~\ref{tab:detection_limit_galaxy}. At {\it g} band, the completeness magnitude yields 24.4\,mag in the wide field, 25.9\,mag in the deep field, and 27.1\,mag in the extreme field.

\begin{figure*}
	\centering
	\includegraphics[width=2\columnwidth]{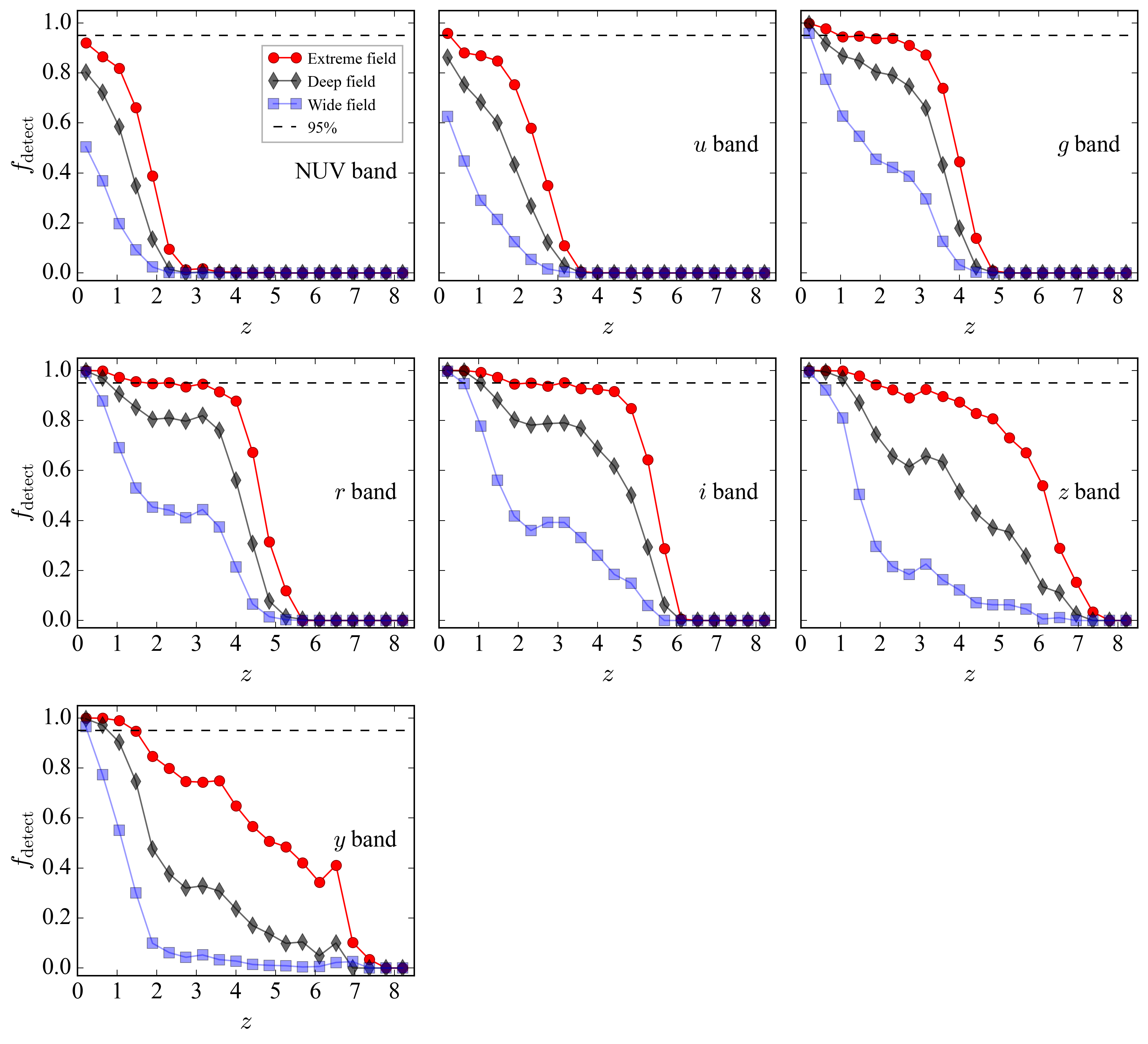}
	\caption{
		Detection completeness ($f_{\mathrm{detect}}$) for galaxies with $M_*>10^9\,M_\odot$ as a function of redshift ($z$) in different bands. 
		The panel layout and plotting conventions are the same as in Fig.~\ref{fig:fp_m}.
	}
	\label{fig:f_z}
\end{figure*}

Figure~\ref{fig:f_z} presents the galaxy ($M_*>10^{9}\,M_\odot$) detection completeness as a function of redshift. In all bands, the detection completeness decreases toward higher redshift, reflecting the increasing difficulty of detecting galaxies at earlier cosmic times. This trend arises because galaxies become progressively fainter and more susceptible to noise at higher redshift, primarily due to cosmological surface-brightness dimming, bandpass shifting, and the redshift evolution of galaxy sizes \cite{Bouwens2004, Daddi2005, Trujillo2007, Buitrago2008, Oesch2010, Mosleh2012, vanderWel2014, Whitney2019, Allen2025, Yu2026}, as well as the cosmic evolution of galaxy surface brightness \cite{Schade1995, Schade1996, Lilly1998, Roche1998, Labbe2003, Barden2005, Sobral2013, Whitney2020}. A detailed discussion of the formalism describing the impact of these effects on observed surface brightness is provided in Section~3.1 of Ref. \cite{Yu2023}. Deeper survey strategies significantly extend the redshift range over which galaxies can be detected with high completeness. For galaxies with $M_* > 10^9\,M_\odot$, the detection completeness in the wide and deep fields drops below unity at $z \lesssim 1$, whereas the extreme field maintains a completeness level of $\gtrsim 0.95$ out to $z \approx 3$--4.  For comparison, the COSMOS field reaches a 5\,$\sigma$ point-source detection limiting magnitude of 27.2\,mag in F814W \cite{Scoville2007, Koekemoer2007}, which is $\sim$0.8\,mag shallower than that of the CSST extreme field. As a result, COSMOS is complete down to stellar masses of $10^9\,M_\odot$ only up to $z\approx2.8$ \cite{Weaver2022}. The corresponding limiting redshift is therefore slightly lower than that of the CSST extreme field, consistent with the shallower depth of COSMOS. Figure~\ref{fig:f_z} further shows that, at wavelengths bluer than the {\it i} band, galaxies are generally not detected beyond $z \gtrsim 6$, while in the {\it z} band and, in particular, the {\it y} band, a small fraction of galaxies remains detectable at these redshifts.  We note that the EGG mock galaxy catalog is based on observations of galaxies prior to the James Webb Space Telescope. Therefore, other populations such as host galaxies of AGNs or little red dots are not included in this study. 

\begin{table}[H]
	\centering
	\caption{Point-source completeness magnitude, corresponding to the 95\% point-source detection limits, for different CSST survey depths.}
	\label{tab:detection_limit}
	\begin{tabular}{cccc}
		\hline
		Filter & Wide field & Deep field & Extreme field \\
		\hline
		NUV & 25.6 & 26.8 & 28.0 \\
		{\it u} & 25.6 & 26.8 & 27.9 \\
		{\it g} & 26.3 & 27.4 & 28.5 \\
		{\it r} & 26.0 & 27.2 & 28.2 \\
		{\it i} & 25.8 & 26.9 & 28.0 \\
		{\it z} & 25.2 & 26.4 & 27.5 \\
		{\it y} & 24.4 & 25.6 & 26.7 \\
		\hline
	\end{tabular}
\end{table}

\begin{table}[H]
	\centering
	\caption{Galaxy completeness magnitudes corresponding to the 95\% detection limits for galaxies with $M_*>10^9\,M_\odot$ in different CSST survey depths.}
	\label{tab:detection_limit_galaxy}
	\begin{tabular}{cccc}
		\hline
		Filter & Wide field & Deep field & Extreme field \\
		\hline
		NUV      & 22.6 & 24.7 & 26.2 \\
		{\it u}  & 22.5 & 24.9 & 26.3 \\
		{\it g}  & 24.4 & 25.9 & 27.1 \\
		{\it r}  & 24.1 & 25.6 & 26.8 \\
		{\it i}  & 24.0 & 25.5 & 26.7 \\
		{\it z}  & 23.5 & 24.9 & 26.2 \\
		{\it y}  & 22.4 & 24.0 & 25.3 \\
		\hline
	\end{tabular}
\end{table}

As discussed previously, our analysis is based on the EGG mock catalog, which is constructed from HST/F160W-selected samples. 
At $z \gtrsim 2$, samples selected based on HST F160W become increasingly incomplete due to significant selection biases. These biases affect populations recently identified by JWST/NIRCAM, such as the ``Little Red Dots'' (e.g., \cite{perez2023}), and result in galaxies that are HST-dark or HST-faint being systematically missed, including heavily dust-obscured submillimeter galaxies (e.g., \cite{McKinney2025, Ren2025}) and potentially high-redshift quiescent galaxies.
As a consequence, galaxies that are intrinsically more difficult to detect may be underrepresented, which can in turn lead to an overestimation of the detection completeness, especially at high redshift.
Furthermore, some high-redshift star-forming galaxies exhibit clumpy structures, where light is concentrated in small bright regions.
These clumps are particularly prominent in the ultraviolet, and because CSST's optical bands correspond to the rest-frame UV at these redshifts, galaxies hosting such clumps may be more readily detected. Galaxies with bright clumps generally exhibit higher peak surface brightness, further enhancing their detectability compared to pure disks. 
In addition, the unprecedentedly deep and wide-area NUV imaging surveys provide the opportunity to detect sources that were previously unpredictable, suggesting that the current mock catalog may underestimate the detection rate in this band.
However, at lower redshift ($z \lesssim 1$), where the parent samples are more complete and less affected by these selection effects, the impact of this limitation is expected to be significantly reduced, and our results remain broadly reliable. We therefore expect our main conclusions to be most robust within this redshift regime.
A fully self-consistent treatment of these missing populations would require constructing a more complete mock catalog incorporating additional observational constraints, which is beyond the scope of the present study but represents an interesting avenue for future investigation.

\subsection{Robustness of Galaxy Flux and Morphological Measurements} \label{sec:morph}
In this subsection, we quantify the robustness of galaxy flux and morphological measurements by comparing the best-fit parameters derived from mock images with the true input values of the simulated galaxies.
Using different fitting software, we examine how the differences between measured and input parameters depend on the input apparent magnitude, for different bands and CSST/SC survey strategies.
The analysis considers several parameters related to galaxy flux and morphology, including $m$ (Section~\ref{subsubsec:mag}), $R_e$ (Section~\ref{subsubsec:re}), $\mu_e$ (Section~\ref{subsubsec:mue}), $n$ (Section~\ref{subsubsec:sersic}), and $q$ (Section~\ref{subsubsec:axis}).
Throughout this section, quantities with the subscript ``fit'' denote values derived from Sérsic model fitting (e.g., the best-fit apparent magnitude $m_{\mathrm{fit}}$), whereas quantities without this subscript refer to the intrinsic input parameters (e.g., the intrinsic apparent magnitude $m$). 

We use $m$ as the horizontal axis to examine how the best-fit values deviate from their intrinsic values. The magnitude is one of the most fundamental properties of a source in an image and provides a direct, though not exact, proxy for how well the galaxy can be modeled. Brighter galaxies (smaller $m$) generally have higher signal-to-noise ratios and better-resolved structures, and are therefore expected to yield more reliable parameter recovery.

\subsubsection{Magnitude} \label{subsubsec:mag}
\begin{figure*}
	\centering
	\includegraphics[width=1.6\columnwidth]{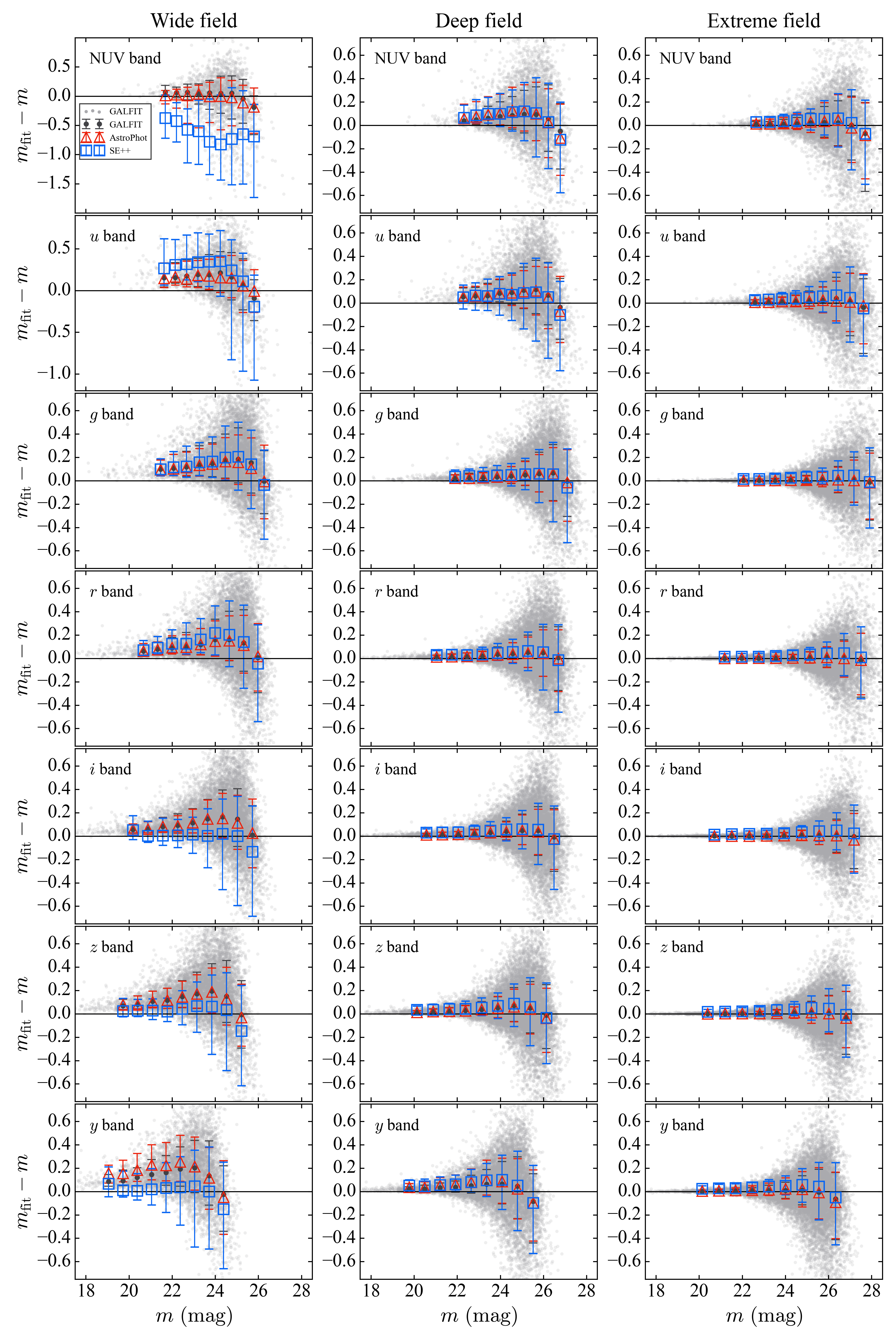}
	\caption{
		Evaluation of biases and uncertainties in apparent magnitude measurements. 
		Each panel shows how the fitted magnitude ($m_{\mathrm{fit}}$) differs from the intrinsic magnitude ($m$) as a function of $m$. 
		Columns correspond to different survey strategies (wide, deep, and extreme), and rows correspond to bands from NUV to {\it y}.
		The background gray points show individual measurement obtained from {\tt GALFIT}.
		The symbols and error bars summarize the median difference obtained with three fitting codes ({\tt GALFIT}, {\tt AstroPhot}, and {\tt SourceXtractor++}),
		with error bars indicating the 16th and 84th percentiles of the distribution in bins of $m$.
		The horizontal solid line marks zero difference for reference.
	}
	
	\label{fig:result_mag}
\end{figure*}

Figure~\ref{fig:result_mag} presents the difference between the measured and input magnitudes, $m_{\mathrm{fit}}-m$, as a function of $m$. 
In each band and for each survey depth, measurements between the 1st and 99th percentiles are divided into nine bins of equal size. For each bin, the median value and the 16th and 84th percentiles are computed and displayed as error bars in Figure~\ref{fig:result_mag}. The gray background points show individual measurements obtained with {\tt GALFIT}. Median values derived from {\tt GALFIT}, {\tt AstroPhot}, and {\tt SourceXtractor++} are indicated by filled black circles, open red triangles, and open blue squares, respectively. In the deep and extreme fields, galaxies show good agreement between $m_{\mathrm{fit}}$ and $m$ over most of the magnitude range, with a small scatter of $\lesssim 0.02$\,mag. Only at the faintest magnitudes does $m_{\mathrm{fit}}$ become underestimated, corresponding to an overestimation of galaxy fluxes, with the scatter increasing to $\approx 0.2$\,mag. This behavior is qualitatively consistent with previous simulation studies of Sloan Digital Sky Survey (SDSS) galaxies \cite{Meert2013} and HST observations \cite{Hiemer2014}. The three codes present consistent results from the deep and extreme fields.

The trend slightly differs in the wide field.
As $m$ increases, these biases generally become slightly larger, suggesting increasing flux underestimation. At the faintest magnitudes, the median offsets decrease and in some cases approach zero or become negative. Overall, for sources brighter than $\sim$ 25 mag, most bands in the deep field yield reasonably accurate magnitude measurements. For fainter sources, the extreme field is required, where the median bias remains close to zero, although the scatter remains substantial at the faint end.
In the NUV and {\it u} bands, {\tt GALFIT} and {\tt AstroPhot} outperform {\tt SourceXtractor++}, with smaller absolute biases (by $\sim$0.17 dex on average and up to $\sim$0.7 dex in the NUV), indicating that the fitting procedure in {\tt SourceXtractor++} is more sensitive to noise in these bands. Nevertheless, in the {\it z} and {\it y} bands, {\tt SourceXtractor++} shows slightly smaller absolute biases (by $\sim$0.1 dex). Overall, the relative performance of the methods does not follow a simple trend, suggesting a more complex interplay between noise properties and fitting procedures.

The three fitting codes yield consistent results in the deep and extreme fields. In the wide field, however, the results from {\tt SourceXtractor++} differ from those obtained with {\tt GALFIT} and {\tt AstroPhot} by up to $\sim 0.1$\,mag. In particular, in the NUV and {\it u} band, {\tt SourceXtractor++} produces more overestimated $m_{\mathrm{fit}}$ values than the other two codes. In the {\it i}, {\it z}, and especially the {\it y} band, {\tt SourceXtractor++} yields measurements that are closer to the input magnitudes and therefore outperforms {\tt GALFIT} and {\tt AstroPhot}. These results indicate that additional bias corrections will be required to achieve accurate galaxy flux measurements in the wide-field CSST surveys.

In Figure~\ref{fig:result_mag}, relatively bright galaxies in the wide survey still exhibit large values of $m_{\mathrm{fit}} - m$, and $\Delta m$ does not vary monotonically with $m$, which appears somewhat unexpected. 
This behavior suggests that the total magnitude $m$ may not fully capture the factors that affect the measurement bias. For extended sources, the effective S/N also depends on the spatial extent of the light distribution and the background noise level, while the total magnitude $m$ characterizes the total flux of the source and the S/N more directly reflects the average signal per pixel relative to the noise. As a result, galaxies with the same apparent magnitude can have substantially different S/N in surveys with different depths.

\begin{figure*}
	\centering
	\includegraphics[width=1.6\columnwidth]{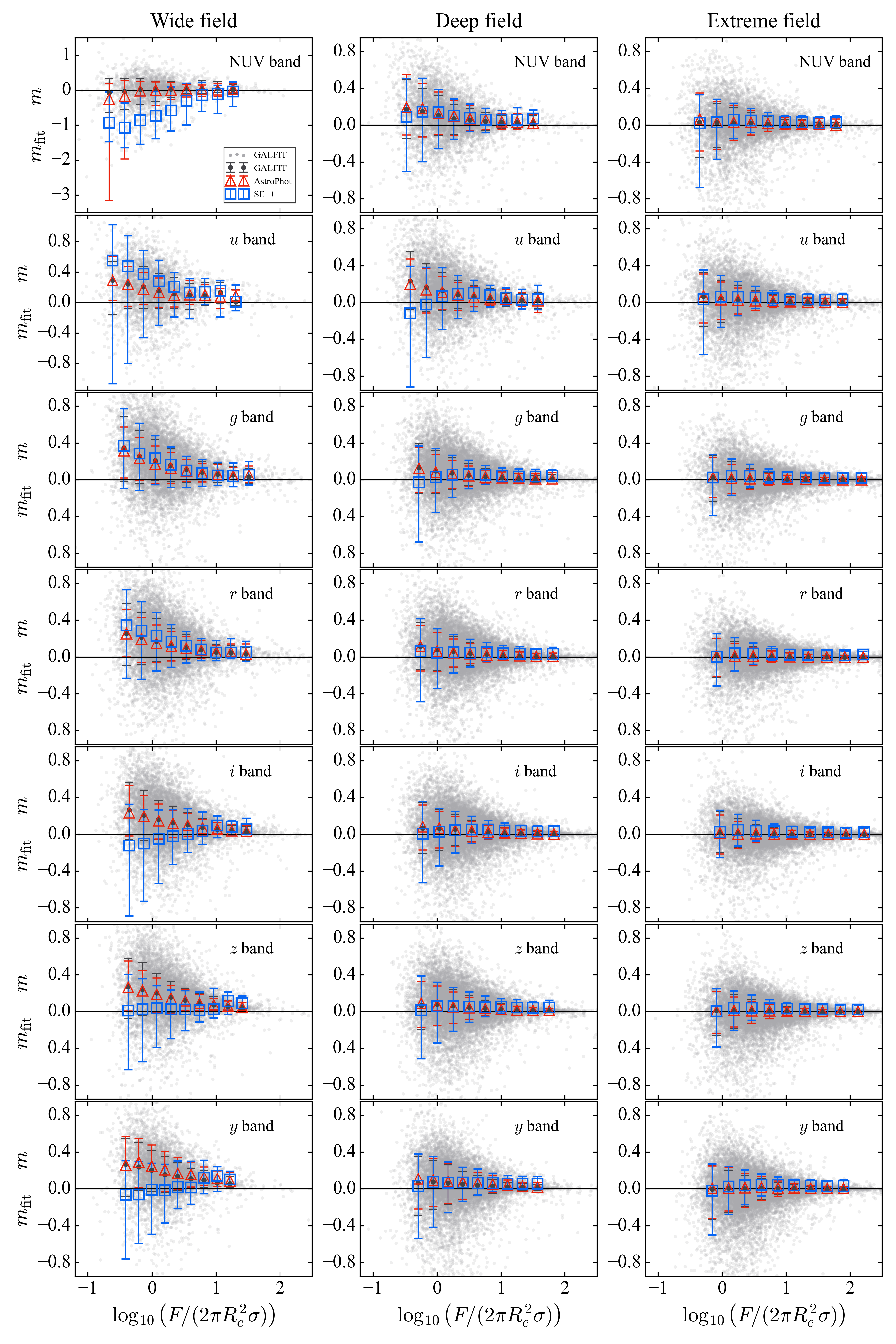}
	\caption{Evaluation of biases and uncertainties in apparent magnitude measurements as a function of a signal-to-noise related parameter. 
	Each panel shows the difference between the fitted magnitude ($m_{\mathrm{fit}}$) and the intrinsic magnitude ($m$) as a function of $\log (F / 2\pi R_e^2 \sigma)$, which approximately characterizes the galaxy signal relative to background noise. 
	Columns correspond to different survey strategies (wide, deep, and extreme), and rows correspond to bands from NUV to {\it y}. 
	The background gray points show individual measurements obtained from {\tt GALFIT}. 
	Symbols and error bars summarize the median difference obtained with three fitting codes ({\tt GALFIT}, {\tt AstroPhot}, and {\tt SourceXtractor++}), with error bars indicating the 16th and 84th percentiles of the distribution in bins of the S/N-related parameter. 
	The horizontal solid line marks zero difference for reference.
	}
	\label{fig:sn}
\end{figure*}

To more directly compare the measurement biases under similar observational conditions, we define an approximate S/N quantity per pixel: 
\begin{equation}\label{SNR}
  \log(\mathrm{S/N}) = \log\left(\frac{F}{2\pi R_e^{2}\sigma}\right),
\end{equation}
where $F$ is the total flux of the galaxy, $R_e$ is the effective radius, and $\sigma$ is the background noise level. It approximately characterizes the signal per unit area relative to the background noise and can be used to compare measurement performance across surveys of different depths.
As shown in Figure~\ref{fig:sn}, when plotted as a function of this parameter, the magnitude bias $\Delta m$ decreases with increasing S/N, and this trend is consistent across the wide, deep, and extreme surveys. This indicates that the apparent differences seen in Figure~\ref{fig:result_mag} mainly arise from the fact that galaxies with the same magnitude do not necessarily have the same S/N in surveys with different depths, rather than from additional systematic biases in the measurement methods themselves.

\begin{figure*}
	\centering
	\includegraphics[width=1.6\columnwidth]{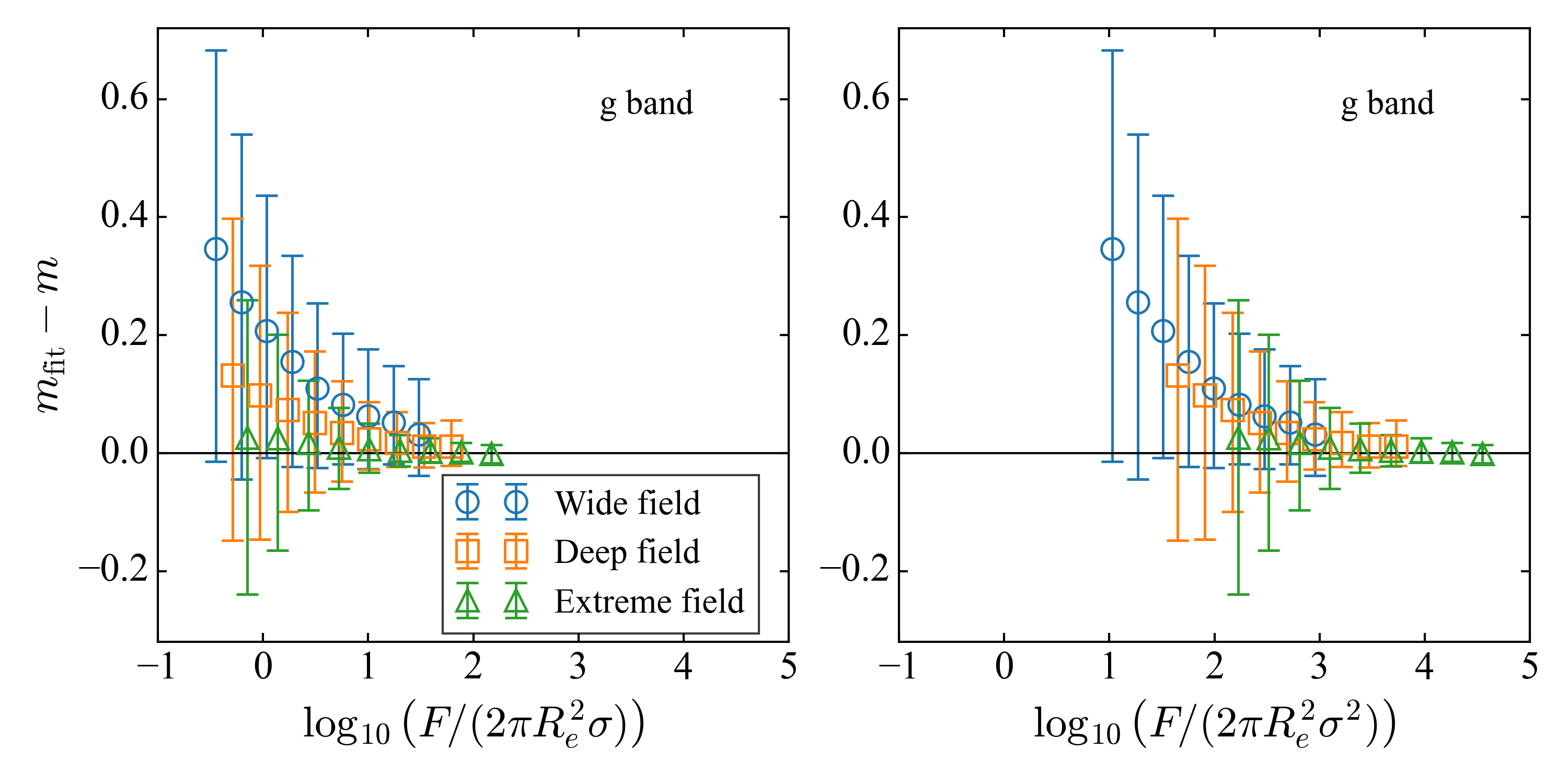}
	\caption{Comparison of magnitude biases as a function of two definitions of an S/N-related parameter in the $g$ band. 
	The left panel shows $m_{\mathrm{fit}} - m$ as a function of $\log (F / 2\pi R_e^2 \sigma)$, while the right panel uses $\log (F / 2\pi R_e^2 \sigma^2)$. 
	Different colors represent the three survey strategies (wide, deep, and extreme). 
	Points indicate the median difference measured with {\tt GALFIT} in bins of the horizontal parameter, with error bars corresponding to the 16th and 84th percentiles of the distribution. 
	The horizontal solid line marks zero difference for reference.}
	\label{fig:fig_gband_compare}
\end{figure*}

However, at a fixed value of $\log (F / 2\pi R_e^2 \sigma)$, the median $m_{\mathrm{fit}} - m$ in the wide survey remains larger than that in the deep survey, suggesting that the adopted S/N definition does not fully capture all factors influencing the magnitude bias. One possible explanation is that this definition underestimates the impact of background noise.
To account for this, we introduce an alternative indicator that increases the weighting of the noise term, defined as $\log (F / 2\pi R_e^2 \sigma^2)$. Figure~\ref{fig:fig_gband_compare} shows $m_{\mathrm{fit}} - m$ as a function of $\log (F / 2\pi R_e^2 \sigma)$ (left) and this alternative indicator (right) for the {\it g}-band simulated images.
Using this indicator, the difference in the median $m_{\mathrm{fit}} - m$ at a fixed horizontal-axis value is reduced, while the overall trend of the magnitude bias is preserved. This may support our interpretation that background noise plays a more significant role than captured by the original S/N definition.

\subsubsection{Effective Radius} \label{subsubsec:re}
\begin{figure*}
	\centering
	\includegraphics[width=1.6\columnwidth]{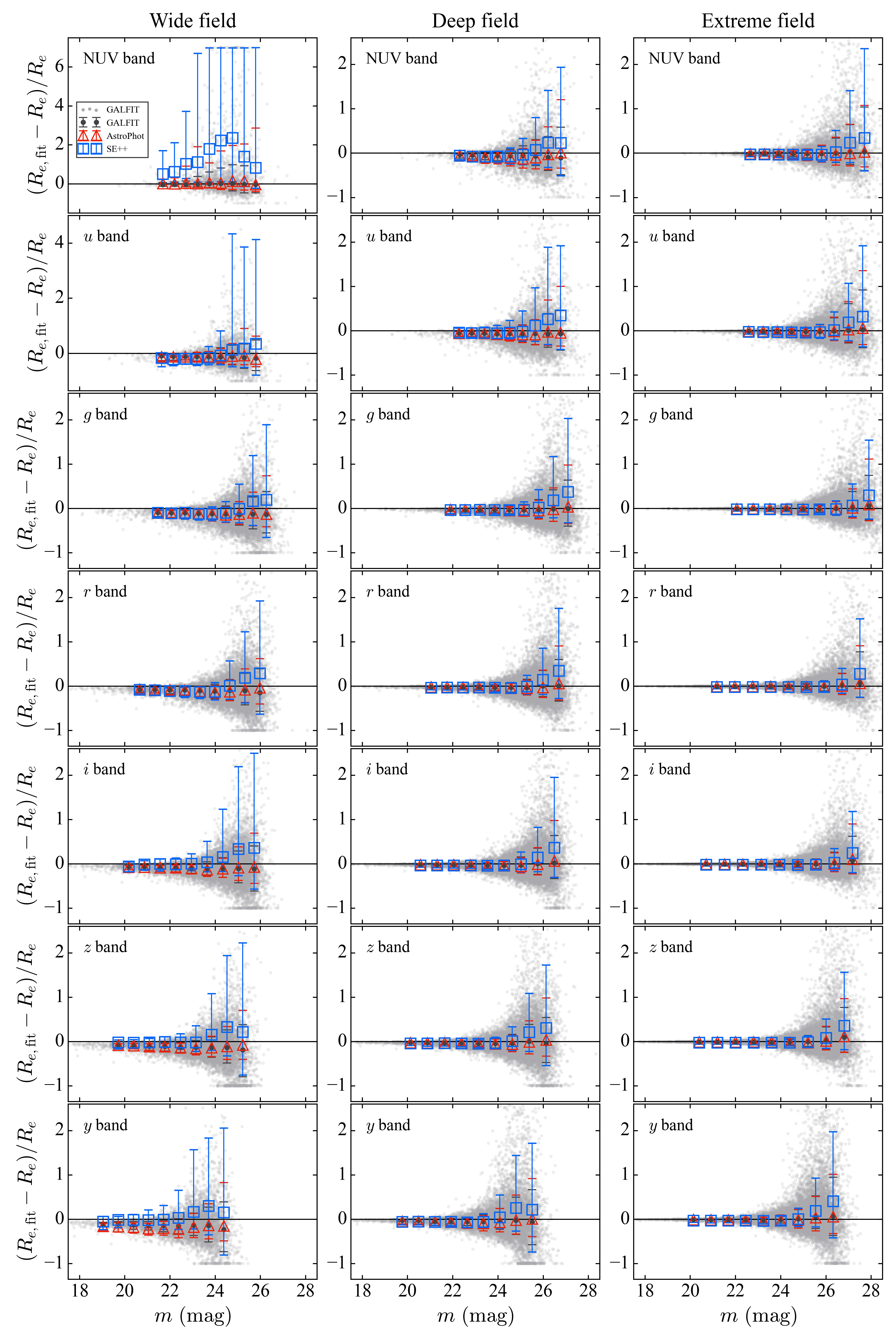}
	\caption{
		Evaluation of biases and uncertainties in effective radius ($R_e$) measurements. 
		Each panel shows how the fitted effective radius ($R_{e,\mathrm{fit}}$) differs from the input value ($R_e$), expressed as the relative difference $(R_{e,\mathrm{fit}} - R_e)/R_e$, as a function of apparent magnitude $m$. 
		The panel layout and plotting conventions are the same as in Fig.~\ref{fig:result_mag}.
	}
	\label{fig:result_Re}
\end{figure*}

Figure~\ref{fig:result_Re} presents the fractional difference of the best-fit effective radius from the input value, $(R_{e,\mathrm{fit}} - R_e)/R_e$, as a function of intrinsic apparent magnitude $m$. 
For bright galaxies, $R_{e,\mathrm{fit}}$ is generally consistent with the input $R_e$ across most bands in the wide, deep, and extreme fields, except for the results obtained with {\tt SourceXtractor++} in the NUV wide field. In these regimes, the median relative deviation is close to zero, at the level of approximately 2\% in the deep field and approximately 0.3\% in the extreme field, with a negligible scatter of approximately 1\%.

As galaxies become fainter, the median fractional difference $(R_{e,\mathrm{fit}} - R_e)/R_e$, derived using {\tt GALFIT} and {\tt AstroPhot}, remains close to zero, although the scatter increases substantially, reaching values of up to approximately 50\%. This behavior is consistent with previous simulation studies based on SDSS data, which show that for bright galaxies, single-Sérsic fitting can recover effective radii with a scatter of $5$\% \cite{Meert2013}. The increasing scatter with fainter $m$ is consistent with the trend reported in Ref. \cite{Euclid2023XXVI}, where Sérsic fits to simulated Euclid images exhibit a characteristic trumpet-shaped distribution, with increasing scatter toward fainter magnitudes. The ability to recover effective radii through Sérsic fitting has also been demonstrated using simulated high-redshift JWST images constructed from real nearby galaxies \cite{Yu2023}.

Overall, survey depth primarily affects the scatter. For most bands, the median fractional differences of effective radius stay close to zero at all depths, whereas deeper surveys show markedly smaller scatter.

Finally, we compare the performance of different fitting codes. {\tt GALFIT} and {\tt AstroPhot} generally outperform {\tt SourceXtractor++} in recovering effective radii. Although all three codes show comparable performance at the bright end, {\tt SourceXtractor++} tends to overestimate effective radii and exhibits larger scatter at fainter magnitudes, especially in the NUV band, where the median overestimation can reach up to $\sim$200\%. An exception is found in the {\it y} band of the wide field, where {\tt SourceXtractor++} produces smaller biases than the other two codes.

Unlike for magnitude bias, using S/N (Equation~(\ref{SNR})) rather than $m$ as the horizontal axis does not reduce the bias in effective radius, nor, as shown below, in effective surface brightness, Sérsic index, or axis ratio.

\subsubsection{Effective Surface Brightness} \label{subsubsec:mue}

\begin{figure*}
	\centering
	\includegraphics[width=1.6\columnwidth]{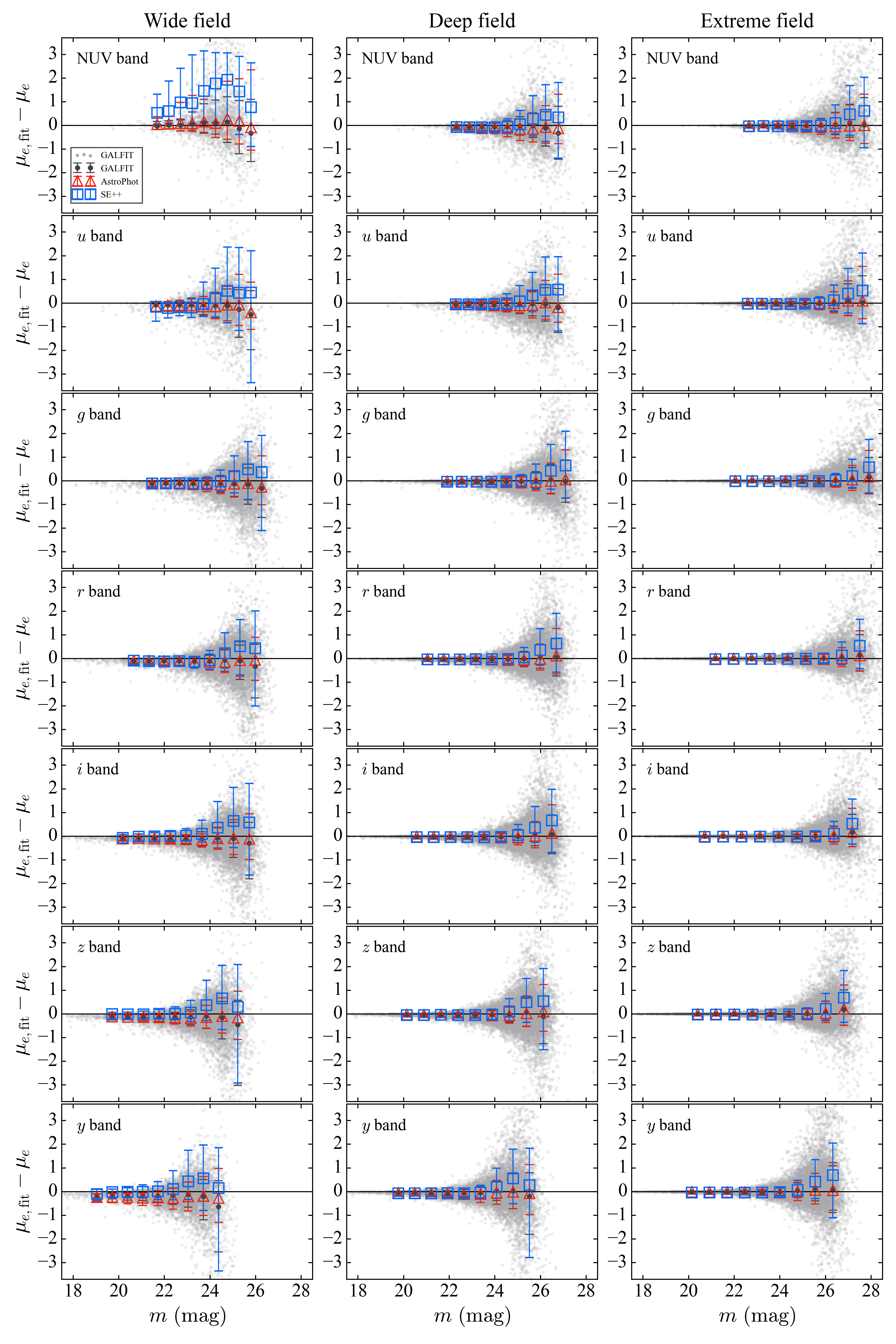}
	\caption{
		Evaluation of differences and uncertainties in effective surface brightness ($\mu_e$) measurements. 
		Each panel shows how the fitted effective surface brightness ($\mu_{e,\mathrm{fit}}$) differs from the input value ($\mu_e$) as a function of apparent magnitude $m$.  
		The panel layout and plotting conventions are the same as in Fig.~\ref{fig:result_mag}.
	}
	\label{fig:result_mu}
\end{figure*}

Figure~\ref{fig:result_mu} shows the deviation of the best-fit effective surface brightness from the input value, $\mu_{e,\mathrm{fit}}-\mu_e$, as a function of apparent magnitude $m$. By definition, $\mu_e$ depends explicitly on both $m$ and $R_e$, and therefore any bias in $\mu_e$ is fully determined by the measurement biases in these two quantities. As illustrated in Figure~\ref{fig:result_mu}, the behavior of $\mu_{e,\mathrm{fit}}-\mu_e$ closely follows that of $(R_{e,\mathrm{fit}} - R_e)/R_e$, due to the logarithmic dependence of $\mu_e$ on $R_e^2$.

Consistent with the measurement biases of $R_e$, in the NUV band of the wide field, the effective surface brightness $\mu_e$ measured with {\tt SourceXtractor++} is significantly overestimated. Apart from this case, the values of $\mu_{e,\mathrm{fit}}-\mu_e$ obtained with {\tt GALFIT} and {\tt AstroPhot} remain close to zero, with the scatter increasing toward fainter magnitudes. In contrast, the measurements from {\tt SourceXtractor++} show systematic overestimation at the faint end, reaching approximately 0.5\,dex.

In the deep and extreme fields, the $\mu_e$ measurements are substantially improved, exhibiting both smaller biases and reduced scatter across all bands. This improvement is primarily driven by the higher signal-to-noise ratios achieved in these deeper surveys.

\subsubsection{Sérsic Index} \label{subsubsec:sersic}
\begin{figure*}
	\centering
	\includegraphics[width=1.6\columnwidth]{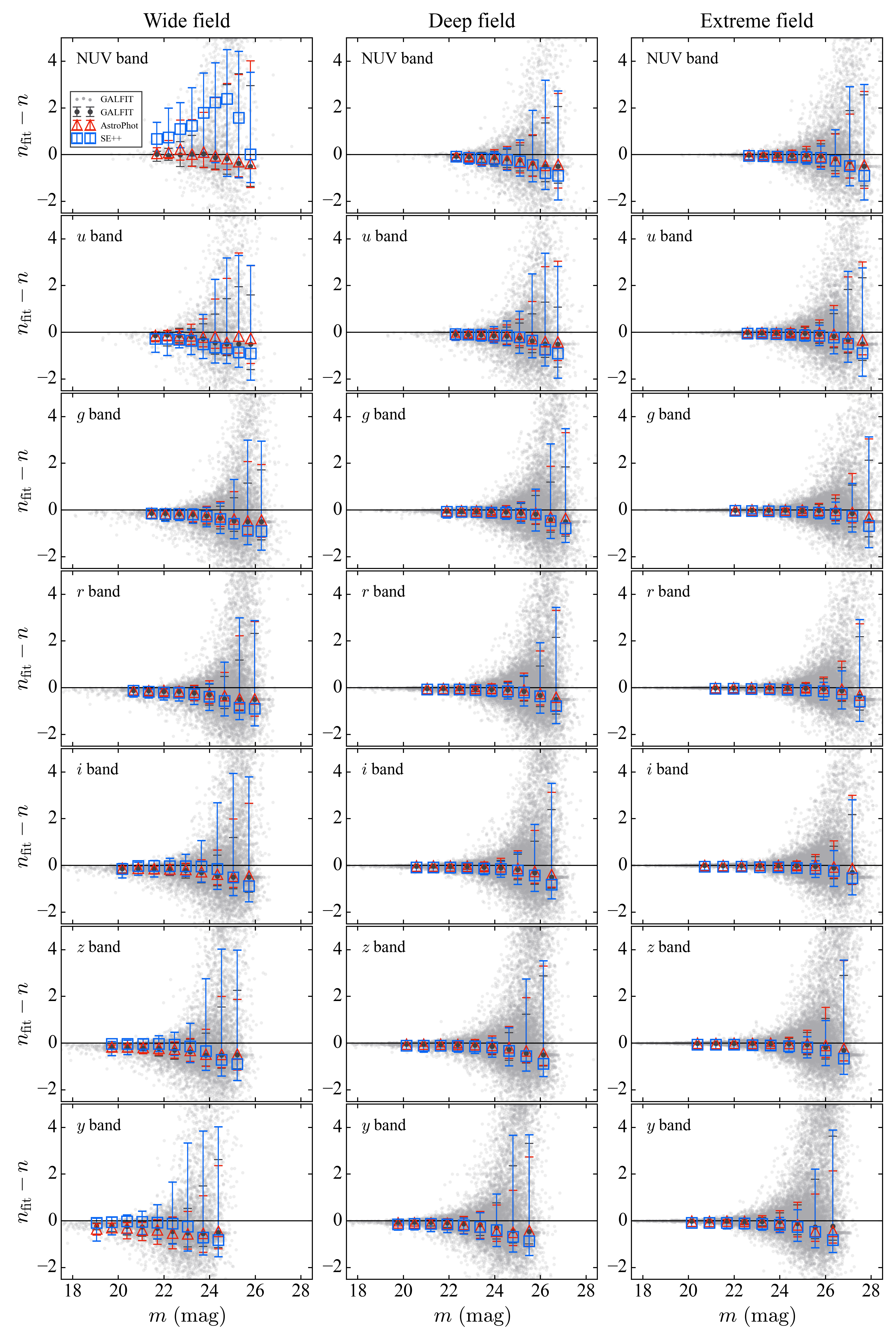}
	\caption{
		Evaluation of differences and uncertainties in Sérsic index ($n$) measurements. 
		Each panel shows how the fitted Sérsic index ($n_{\mathrm{fit}}$) differs from the input value ($n$) as a function of apparent magnitude $m$. 
		The panel layout and plotting conventions are the same as in Fig.~\ref{fig:result_mag}.
	}
	\label{fig:result_n}
\end{figure*}

Figure~\ref{fig:result_n} presents the deviation of the best-fit Sérsic index $n_{\mathrm{fit}}$ from the input value $n$ ($n_{\mathrm{fit}}-n$) as a function of apparent magnitude $m$. For bright galaxies, the median values of $n_{\mathrm{fit}}-n$ obtained with {\tt GALFIT} and {\tt AstroPhot} are close to zero with negligible scatter across most bands. 
An exception occurs in the wide-field NUV band measured with {\tt SourceXtractor++}, where $n_{\mathrm{fit}}$ is significantly overestimated and exhibits substantial scatter. In addition, the {\it u} band in the wide field shows slightly larger deviations at the bright end compared to other bands, likely due to their relatively low signal-to-noise ratios. 
These behaviors are consistent with previous simulation studies that focus on bright galaxies \cite{Meert2013,Hiemer2014,Euclid2023XXVI}.

As galaxies become fainter, the median values of $n_{\mathrm{fit}}-n$ shift mildly toward negative values, reaching $\sim\,-0.4$ at the faintest magnitudes. At the same time, the scatter changes markedly. The 16th percentile of $n_{\mathrm{fit}}-n$ rises modestly from $\sim 0.05$ at the bright end to $\sim 0.5$ at the faint end, while the 84th percentile increases dramatically from $\sim 0.03$ to values as large as $\sim 4$. This significant asymmetry indicates that, for faint galaxies, a substantial fraction of measurements severely overestimate the Sérsic index. As a result, classifications that rely on $n$ to distinguish between disk-dominated and spheroid-dominated systems should be treated with caution in this regime.

The dependence of Sérsic index measurements on survey depth is modest. Deeper exposures reduce the scatter, but the median trends remain broadly similar across the wide, deep, and extreme fields. Results are generally consistent across different bands. {\tt GALFIT} and {\tt AstroPhot} yield mutually consistent results. In comparison, {\tt SourceXtractor++} shows slightly poorer performance at faint magnitudes, with $n_{\mathrm{fit}}$ being more strongly underestimated on average and exhibiting larger uncertainties.

\subsubsection{Axis Ratio} \label{subsubsec:axis}
\begin{figure*}
	\centering
	\includegraphics[width=1.6\columnwidth]{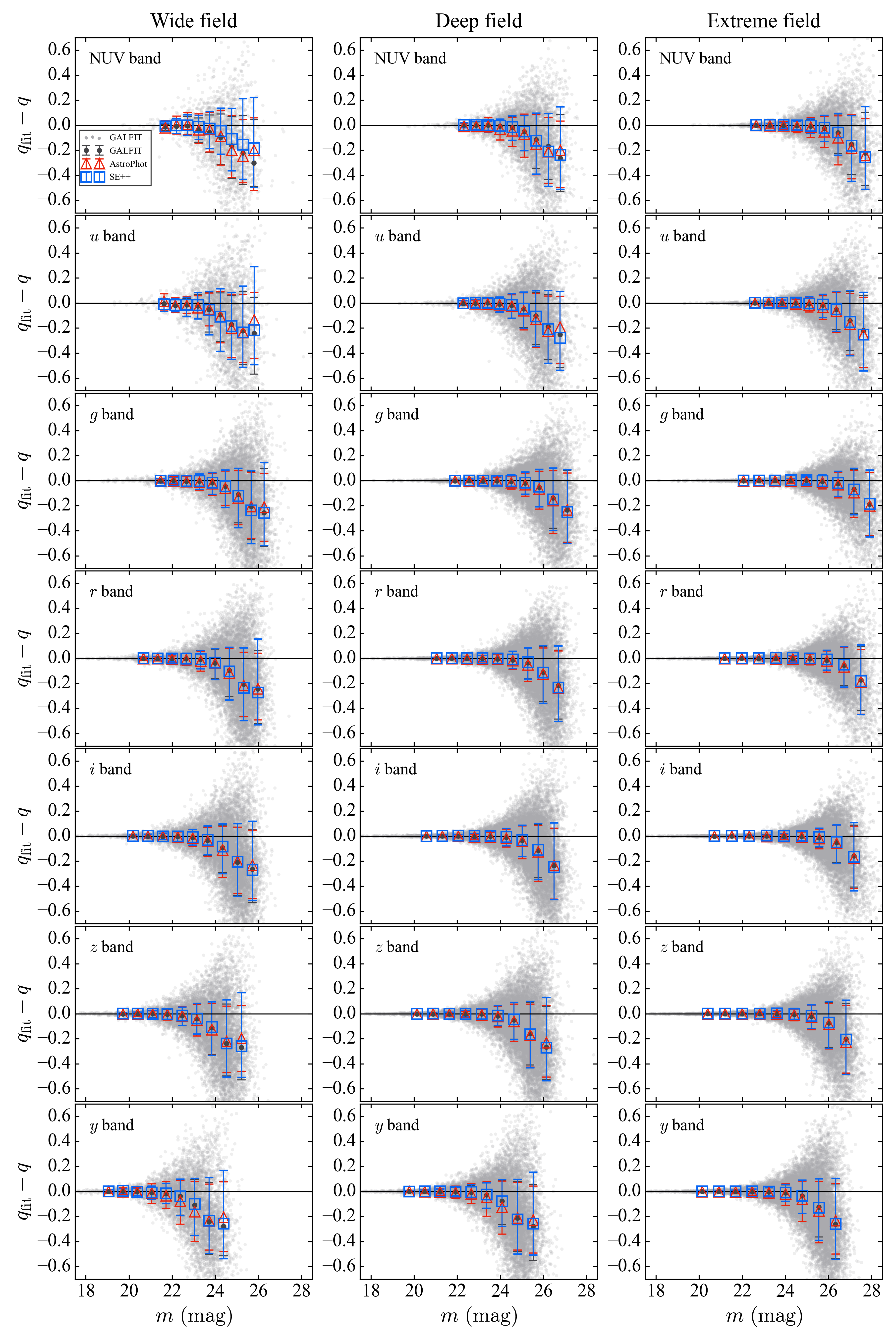}
	\caption{
		Evaluation of differences and uncertainties in axis ratio ($q$) measurements. 
		Each panel shows how the fitted axis ratio ($q_{\mathrm{fit}}$) differs from the input value ($q$) as a function of apparent magnitude $m$. 
		The panel layout and plotting conventions are the same as in Fig.~\ref{fig:result_mag}.
	}
	\label{fig:result_q}
\end{figure*}

Figure~\ref{fig:result_q} shows the deviation of the best-fit axis ratio $q_{\mathrm{fit}}$ from the input value $q$ ($q_{\mathrm{fit}}-q$) as a function of apparent magnitude $m$. For bright galaxies, $q_{\mathrm{fit}}$ closely matches the input $q$ with negligible scatter across all bands and survey depths. As galaxies become fainter, $q_{\mathrm{fit}}$ is progressively underestimated and the associated scatter increases. At the faintest magnitudes, $q_{\mathrm{fit}}$ underestimates $q$ by up to $\sim\,0.3$. This systematic decrease in $q_{\mathrm{fit}}-q$ is observed for all survey depths, although the magnitude at which the bias becomes significant depends on depth: at $m \sim 22$ in the wide field, $m \sim 23$ in the deep field, and $m \sim 25$ in the extreme field.

These results are broadly consistent with Ref. \cite{Euclid2023XXVI}, who showed that axis ratios for simulated bright Euclid galaxies are generally recovered with high accuracy across different fitting methods, exhibiting negligible systematic bias and low dispersion over a wide magnitude range in the optical {\it I} band. Differences among bands are modest, with similar behavior observed in most cases. The three fitting codes, {\tt GALFIT}, {\tt AstroPhot}, and {\tt SourceXtractor++}, yield mutually consistent results.

\subsubsection{Limitations of the adopted Sérsic Models}
In this study, we adopt single-component Sérsic models based on the assumption that pure disks and bulges can be approximated by Sérsic profiles with $n=1$ and $n=4$, respectively. 
Observations in the rest-frame optical indicate that a significant fraction of star-forming galaxies at $z \sim 1$--3 exhibit clumpy morphologies with $n$ not equal 1 (e.g., \cite{Guo2015,Bournaud2007,Yu2025}). 
High-redshift quiescent galaxies similarly tend to have Sérsic indices below $n=4$ (e.g., \cite{Martorano2025}), indicating that the fixed $n=4$ assumption does not fully capture their structural diversity.
Therefore, the recovery of Sérsic indices across the entire galaxy sample may be biased, as the simplified input models do not fully account for the diverse and complex morphologies of high-redshift galaxies.

\subsection{Future Prospects}

The unprecedented volume and depth of imaging data delivered by CSST will enable transformative studies of galaxy formation and evolution across a wide range of cosmic environments and epochs. In this context, understanding systematic biases in the measurement of galaxy fluxes and structural parameters is essential for fully exploiting the scientific potential of the survey.

Our results demonstrate that measurement biases, particularly for faint sources, can lead to systematic underestimation of galaxy fluxes and distortions in recovered structural parameters. 
Once galaxies are sufficiently resolved, these biases are primarily driven by low S/N and parameter degeneracies.
Such biases directly propagate into derived physical quantities, including stellar masses and star formation rates obtained through spectral energy distribution fitting (e.g., using tools such as CIGALE\cite{Boquien2019}). Quantifying and correcting these effects will therefore be critical for ensuring the accuracy of galaxy population studies. A natural extension of this work is to directly assess how observational biases at the image level impact inferred physical properties, by combining realistic image simulations with SED modeling.

Accurate structural measurements are also essential for studies linking morphology to galaxy evolution. For example, investigations of environmental effects on galaxy morphology, the role of structure in star formation quenching, and the impact of mergers on galaxy evolution all rely on robust measurements of parameters such as Sérsic index and effective radius. Systematic biases in these quantities may lead to misinterpretation of morphological trends or evolutionary pathways. The bias calibrations presented in this work provide a framework to both qualitatively and quantitatively account for such effects, enabling more reliable comparisons between observations and theoretical models.

Future work may extend the present analysis by incorporating additional observational complexities, including source blending, spatially varying backgrounds, and crowded environments. In particular, a comprehensive treatment of low surface brightness galaxies and multi-component structural fitting will be important for capturing the full diversity of galaxy populations in CSST data. Together, these efforts will further improve the robustness of structural measurements and maximize the scientific return of CSST.

\section{Conclusion}  \label{sec:conclusion}
The upcoming CSST is poised to deliver high-resolution imaging of a vast population of galaxies from the NUV to the y band. To evaluate CSST's photometric performance and support early-science analysis, we simulated the telescope's Wide, Deep, and Extreme Deep survey modes (exposures of 300\,s, 2000\,s, and 15,000\,s, respectively). To characterize CSST's galaxy detection and photometric capabilities in preparation for early-science programs, we generated 470,526 mock images for 22,406 simulated galaxies with $M_*>10^9\,M_\odot$ across all seven filters and three survey depths.  We then performed source detection and photometric analysis using three independent fitting codes: {\tt GALFIT}, {\tt AstroPhot}, and {\tt SourceXtractor++}.  By comparing these results, we quantified the measurement biases and uncertainties in galaxy magnitude ($m$), effective radius ($R_e$), effective surface brightness ($\mu_e$), S\'ersic index ($n$), and axis ratio ($q$).   We note that all simulations in this work assume isolated galaxies, and the results thus represent an optimistic, effectively upper limit estimate of the detection and measurement performance achievable with CSST imaging under idealized conditions.  Our main results are summarized as follows.

\begin{enumerate}
	\item For point sources, the 95\% completeness magnitude in the {\it g} band reaches 26.3, 27.4, and 28.5\,mag in the wide, deep, and extreme fields, respectively. The corresponding limits in all bands are listed in Table~\ref{tab:detection_limit}.
	
	\item For galaxies ($M_*>10^9\,M_\odot$), the 95\% completeness magnitudes in the {\it g} band are 24.4, 25.9, and 27.1\,mag for the wide, deep, and extreme fields, respectively. The corresponding limits in all bands are listed in Table~\ref{tab:detection_limit_galaxy}. A detection completeness above 95\% is maintained at $z=3$--$4$ only in the extreme field, while the corresponding redshift limits are $z\approx1$ in the deep field and $z\approx0.5$ in the wide field.
	
	\item On average, for fainter galaxies, the reduced signal-to-noise ratio leads to systematic overestimates in $m$, $R_e$, and $\mu_e$, and underestimates in $n$ and $q$. These biases, as well as the associated scatter, become progressively smaller in deeper fields, as quantified in Figures~\ref{fig:result_mag}--\ref{fig:result_q}. 
 		
	\item In most bands and across the Wide, Deep, and Extreme fields, {\tt GALFIT}, {\tt AstroPhot}, and {\tt SourceXtractor++} deliver consistent results, particularly for bright galaxies. For faint galaxies, however, {\tt GALFIT} and {\tt AstroPhot} yield measurements, especially of $R_e$ and $n$, with smaller biases and smaller uncertainties than {\tt SourceXtractor++}.
	
\end{enumerate}

Overall, our results provide quantitative constraints on sample selection and on the robustness of photometric and morphological measurements for both CSST early-science programs and future legacy surveys.

%%%%%%%%%%%%%%%%%%%%%%%%%%%%%%%%%%%%%%%%%%%%%%%%%%%%%%%
%%% Acknowledgements. ??§Ý
%%%%%%%%%%%%%%%%%%%%%%%%%%%%%%%%%%%%%%%%%%%%%%%%%%%%%%%
\Acknowledgements{We thank the referees for the insightful comments and suggestions. This work is supported by the National SKA Program of China No. 2025SKA0150103 and the National Natural Science Foundation of China under Nos. 12550002, 12133008, 12221003, 12233001, and 12533004. We acknowledge the science research grants from the China Manned Space Project with No. CMS-CSST-2021-A04, No. CMS-CSST-2025-A10, and No. CMS-CSST-2025-A09, the National Key R\&D Program of China No. 2024YFA1611602, the Fundamental Research Funds for the Central Universities, Peking University No. 7100604896, and the Shanghai Natural Science Research Grant No. 24ZR1491200.
}

\InterestConflict{The authors declare that they have no conflict of interest.}

%%%%%%%%%%%%%%%%%%%%%%%%%%%%%%%%%%%%%%%%%%%%%%%%%%%%%%%
%%% Conflict of interest. ????????????
%%%%%%%%%%%%%%%%%%%%%%%%%%%%%%%%%%%%%%%%%%%%%%%%%%%%%%%

%%%%%%%%%%%%%%%%%%%%%%%%%%%%%%%%%%%%%%%%%%%%%%%%%%%%%%%
%%% Supplements. ????????, ????
%%%%%%%%%%%%%%%%%%%%%%%%%%%%%%%%%%%%%%%%%%%%%%%%%%%%%%%
%\Supplements{}

%%%%%%%%%%%%%%%%%%%%%%%%%%%%%%%%%%%%%%%%%%%%%%%%%%%%%%%
%%% Reference section. ?¦Ï?????
%%% citation in the content using "some words~\cite{1,2}".
%%% ~ is needed to make the reference number is on the same line with the word before it.
%%%%%%%%%%%%%%%%%%%%%%%%%%%%%%%%%%%%%%%%%%%%%%%%%%%%%%%

\bibliographystyle{scpma}
\bibliography{ref}

\begin{thebibliography}{10}
\providecommand{\url}[1]{\texttt{#1}}
\providecommand{\urlprefix}{URL }
\providecommand{\doi}[1]{doi:~\href{http://doi.org/#1}{\nolinkurl{#1}}}
\providecommand{\arXiv}[1]{\href{https://arxiv.org/abs/#1}{\nolinkurl{https://arxiv.org/abs/#1}}}
\providecommand{\eprint}[1]{\href{http://arxiv.org/abs/#1}{\nolinkurl{#1}}}

\bibitem{Zhan2021}
H.~{Zhan}, Chinese Science Bulletin \textbf{66}, 1290 (2021).

\bibitem{Yan2025IFU}
Z.-J. {Yan}, J.~{Yin}, L.~{Hao}, S.-Y. {Shen}, W.~{Chen}, S.~{Feng}, Y.-F.
  {Xiong}, C.~{Xu}, X.-R. {Wen}, L.~{Lin}, C.~{Liu}, L.~{Long}, Z.-L. {Chen},
  M.-C. {Wu}, X.-B. {Li}, Z.~{Ban}, X.~{Yang}, Y.-X. {Jiang}, G.-L. {Li}, K.-X.
  {Li}, J.-J. {Chen}, N.~{Li}, C.-L. {Wei}, L.~{Wang}, B.-C. {Ren}, J.~{Wei},
  J.~{Tang}, and R.~{Li}, arXiv e-prints arXiv:2511.12483 (2025), arXiv:
  \eprint{2511.12483}.

\bibitem{Yan2025MCI}
Z.-J. {Yan}, H.-Y. {Shan}, Z.-Y. {Zheng}, X.-Y. {Peng}, Z.-X. {Qi}, C.~{Xu},
  L.~{Lin}, X.-R. {Wen}, C.-Y. {Jiang}, L.-X. {Zheng}, J.~{Zhong}, F.-T.
  {Yuan}, Z.-L. {Chen}, W.~{Chen}, M.-C. {Wu}, Z.-S. {Fu}, K.-X. {Li},
  L.~{Nie}, C.~{Liu}, N.~{Li}, Q.~{Wang}, Z.-H. {Cao}, S.~{Feng}, G.-L. {Li},
  L.~{Wang}, C.-L. {Wei}, X.-B. {Li}, Z.~{Ban}, X.~{Yang}, Y.-X. {Jiang}, D.-Z.
  {Liu}, Y.-H. {Chen}, X.-H. {Liu}, F.~{Xu}, X.~{Cheng}, Y.~{Su}, T.-F. {Duan},
  C.~{Qi}, N.~{Li}, G.~{Zheng}, C.~{Ma}, J.~{Tang}, and R.~{Li}, arXiv e-prints
  arXiv:2511.12481 (2025), arXiv: \eprint{2511.12481}.

\bibitem{Wei2025Mock}
C.-L. {Wei}, Y.~{Luo}, H.~{Tian}, M.~{Li}, Y.-S. {Qiu}, G.-L. {Li}, Y.-D.
  {Fang}, X.~{Zhang}, D.-Z. {Liu}, N.~{Li}, R.~{Li}, H.-Y. {Shan}, L.~{Nie},
  Z.~{He}, L.~{Wang}, X.~{Kang}, D.~{Fan}, Y.~{Chen}, X.~{Fu}, and C.~{Liu},
  arXiv e-prints arXiv:2511.10805 (2025), arXiv: \eprint{2511.10805}.

\bibitem{Zhu2025CPIC}
Y.-M. {Zhu}, G.~{Zhao}, J.-P. {Dou}, Z.-H. {Lv}, Y.-L. {Chen}, B.~{Ma}, Z.-J.
  {Yan}, J.~{Tang}, and R.~{Li}, arXiv e-prints arXiv:2511.09862 (2025), arXiv:
  \eprint{2511.09862}.

\bibitem{Tan2025HSTDM}
S.~{Tan}, W.~{Duan}, Y.~{Zhang}, Y.~{Ao}, Y.~{Gong}, Z.~{Lin}, X.~{Zhang},
  Y.~{Shi}, J.~{Tang}, J.~{Li}, R.~{Mao}, and S.-C. {Shi}, arXiv e-prints
  arXiv:2511.09074 (2025), arXiv: \eprint{2511.09074}.

\bibitem{Zhao2025CPIC}
G.~{Zhao}, Y.~{Zhu}, J.~{Dou}, Y.~{Chen}, Z.~{Lv}, B.~{Niu}, Z.~{Yan}, B.~{Ma},
  and R.~{Li}, arXiv e-prints arXiv:2511.08886 (2025), arXiv:
  \eprint{2511.08886}.

\bibitem{Wei2025Overview}
C.-L. {Wei}, G.-L. {Li}, Y.-D. {Fang}, X.~{Zhang}, Y.~{Luo}, H.~{Tian}, D.-Z.
  {Liu}, X.-M. {Meng}, Z.~{Ban}, X.-B. {Li}, Z.~{Luo}, J.-T. {Xian}, W.~{Wang},
  X.-Y. {Peng}, N.~{Li}, R.~{Li}, L.~{Shao}, T.-M. {Zhang}, J.~{Tang},
  Y.~{Chen}, Z.-X. {Qi}, Z.-H. {Cao}, H.-Y. {Shan}, L.~{Nie}, L.~{Wang},
  Z.~{He}, R.-B. {Luo}, Q.-Y. {Liu}, and Z.-J. {Yan}, arXiv e-prints
  arXiv:2511.06970 (2025), arXiv: \eprint{2511.06970}.

\bibitem{Xian2025}
X.~{Jing-Tian}, L.~{Lin}, F.~{Yue-Dong}, Z.~{Xin}, X.~{You-Hua}, M.~{Xian-Min},
  T.~{Hao}, Z.~{Tian-Yi}, B.~{Zhang}, L.~{Guo-Liang}, X.~{Shu-Yan}, and
  W.~{Wei}, arXiv e-prints arXiv:2511.06956 (2025), arXiv: \eprint{2511.06956}.

\bibitem{Ban2025}
Z.~{Ban}, X.-B. {Li}, X.~{Yang}, Y.-X. {Jiang}, H.-C. {Ma}, W.~{Wang}, J.-g.
  {Lv}, C.-L. {Wei}, D.-Z. {Liu}, G.-L. {Li}, C.~{Liu}, N.~{Li}, R.~{Li}, and
  P.~{Wei}, arXiv e-prints arXiv:2511.06936 (2025), arXiv: \eprint{2511.06936}.

\bibitem{Xie2025}
Y.~{Xie}, X.~{Chen}, S.~{Feng}, Z.~{Yan}, N.~{Li}, H.~{Shan}, Y.~{Li},
  C.~{Wei}, W.~{Xu}, Z.~{Zheng}, R.~{Li}, W.~{Chen}, Z.~{Chen}, C.~{Jiang},
  D.~{Liu}, L.~{Nie}, X.~{Peng}, L.~{Wang}, M.~{Wu}, C.~{Xu}, F.~{Yuan},
  S.~{Zhang}, and J.~{Zhong}, arXiv e-prints arXiv:2511.06928 (2025), arXiv:
  \eprint{2511.06928}.

\bibitem{Feng2025}
S.~{Feng}, S.~{Shen}, W.~{Chen}, Z.~{Yan}, R.~{Ye}, J.~{Chen}, X.~{Dai},
  J.~{Ge}, L.~{Hao}, R.~{Li}, Y.~{Liang}, L.~{Lin}, F.~{Liu}, J.~{Lu},
  Z.~{Shao}, M.~{Wu}, Y.~{Xiong}, C.~{Xu}, and J.~{Yin}, arXiv e-prints
  arXiv:2511.06927 (2025), arXiv: \eprint{2511.06927}.

\bibitem{Zhang2025}
X.~{Zhang}, Y.-d. {Fang}, C.-l. {Wei}, G.-l. {Li}, F.-s. {Liu}, H.-x. {Ji},
  H.~{Tian}, N.~{Li}, X.-m. {Meng}, J.-j. {Chen}, X.~{Wang}, R.~{Wang},
  C.~{Liu}, Z.-w. {Hu}, R.~{Li}, P.~{Wei}, and J.~{Tang}, arXiv e-prints
  arXiv:2511.06917 (2025), arXiv: \eprint{2511.06917}.

\bibitem{PracticalStatistics2012}
J.~V. {Wall} and C.~R. {Jenkins}, \emph{{Practical Statistics for Astronomers}}
  (2012).

\bibitem{Shen2003}
S.~{Shen}, H.~J. {Mo}, S.~D.~M. {White}, M.~R. {Blanton}, G.~{Kauffmann},
  W.~{Voges}, J.~{Brinkmann}, and I.~{Csabai}, \mnras \textbf{343}, 978 (2003),
  arXiv: \eprint{astro-ph/0301527}.

\bibitem{Simard2011}
L.~{Simard}, J.~T. {Mendel}, D.~R. {Patton}, S.~L. {Ellison}, and A.~W.
  {McConnachie}, \apjs \textbf{196}, 11 (2011), arXiv: \eprint{1107.1518}.

\bibitem{Shuntov2025}
M.~{Shuntov}, O.~{Ilbert}, S.~{Toft}, R.~C. {Arango-Toro}, H.~B. {Akins}, C.~M.
  {Casey}, M.~{Franco}, S.~{Harish}, J.~S. {Kartaltepe}, A.~M. {Koekemoer},
  H.~J. {McCracken}, L.~{Paquereau}, C.~{Laigle}, M.~{Bethermin}, Y.~{Dubois},
  N.~E. {Drakos}, A.~{Faisst}, G.~{Gozaliasl}, S.~{Gillman}, C.~C. {Hayward},
  M.~{Hirschmann}, M.~{Huertas-Company}, C.~K. {Jespersen}, S.~{Jin},
  V.~{Kokorev}, E.~{Lambrides}, D.~{Le Borgne}, D.~{Liu}, G.~{Magdis},
  R.~{Massey}, C.~J.~R. {McPartland}, W.~{Mercier}, J.~E. {McCleary},
  J.~{McKinney}, P.~A. {Oesch}, A.~{Renzini}, J.~D. {Rhodes}, R.~M. {Rich},
  B.~E. {Robertson}, D.~{Sanders}, M.~{Trebitsch}, L.~{Tresse}, F.~{Valentino},
  A.~P. {Vijayan}, J.~R. {Weaver}, A.~{Weibel}, S.~M. {Wilkins}, and L.~{Yang},
  \aap \textbf{695}, A20 (2025), arXiv: \eprint{2410.08290}.

\bibitem{vanderWel2014}
A.~{van der Wel}, M.~{Franx}, P.~G. {van Dokkum}, R.~E. {Skelton}, I.~G.
  {Momcheva}, K.~E. {Whitaker}, G.~B. {Brammer}, E.~F. {Bell}, H.~W. {Rix},
  S.~{Wuyts}, H.~C. {Ferguson}, B.~P. {Holden}, G.~{Barro}, A.~M. {Koekemoer},
  Y.-Y. {Chang}, E.~J. {McGrath}, B.~{H{\"a}ussler}, A.~{Dekel}, P.~{Behroozi},
  M.~{Fumagalli}, J.~{Leja}, B.~F. {Lundgren}, M.~V. {Maseda}, E.~J. {Nelson},
  D.~A. {Wake}, S.~G. {Patel}, I.~{Labb{\'e}}, S.~M. {Faber}, N.~A. {Grogin},
  and D.~D. {Kocevski}, \apj \textbf{788}, 28 (2014), arXiv:
  \eprint{1404.2844}.

\bibitem{Kartaltepe2023}
J.~S. {Kartaltepe}, C.~{Rose}, B.~N. {Vanderhoof}, E.~J. {McGrath},
  L.~{Costantin}, I.~G. {Cox}, L.~Y.~A. {Yung}, D.~D. {Kocevski}, S.~{Wuyts},
  H.~C. {Ferguson}, M.~B. {Bagley}, S.~L. {Finkelstein}, R.~O. {Amor{\'\i}n},
  B.~H. {Andrews}, P.~A. {Haro}, B.~E. {Backhaus}, P.~{Behroozi},
  L.~{Bisigello}, A.~{Calabr{\`o}}, C.~M. {Casey}, R.~T. {Coogan}, M.~C.
  {Cooper}, D.~{Croton}, A.~{de la Vega}, M.~{Dickinson}, A.~{Fontana},
  M.~{Franco}, A.~{Grazian}, N.~A. {Grogin}, N.~P. {Hathi}, B.~W. {Holwerda},
  M.~{Huertas-Company}, K.~G. {Iyer}, S.~{Jogee}, I.~{Jung}, L.~J. {Kewley},
  A.~{Kirkpatrick}, A.~M. {Koekemoer}, J.~{Liu}, J.~M. {Lotz}, R.~A. {Lucas},
  J.~A. {Newman}, C.~{Pacifici}, V.~{Pandya}, C.~{Papovich}, L.~{Pentericci},
  P.~G. {P{\'e}rez-Gonz{\'a}lez}, J.~{Petersen}, N.~{Pirzkal}, M.~{Rafelski},
  S.~{Ravindranath}, R.~C. {Simons}, G.~F. {Snyder}, R.~S. {Somerville}, E.~R.
  {Stanway}, A.~N. {Straughn}, S.~{Tacchella}, J.~R. {Trump},
  J.~{Vega-Ferrero}, S.~M. {Wilkins}, G.~{Yang}, and J.~A. {Zavala}, \apjl
  \textbf{946}, L15 (2023), arXiv: \eprint{2210.14713}.

\bibitem{Yu2023}
S.-Y. {Yu}, C.~{Cheng}, Y.~{Pan}, F.~{Sun}, and Y.~A. {Li}, \aap \textbf{676},
  A74 (2023), arXiv: \eprint{2307.04753}.

\bibitem{Meert2013}
A.~{Meert}, V.~{Vikram}, and M.~{Bernardi}, \mnras \textbf{433}, 1344 (2013),
  arXiv: \eprint{1211.6123}.

\bibitem{Rowe2015Galsim}
B.~T.~P. {Rowe}, M.~{Jarvis}, R.~{Mandelbaum}, G.~M. {Bernstein}, J.~{Bosch},
  M.~{Simet}, J.~E. {Meyers}, T.~{Kacprzak}, R.~{Nakajima}, J.~{Zuntz},
  H.~{Miyatake}, J.~P. {Dietrich}, R.~{Armstrong}, P.~{Melchior}, and M.~S.~S.
  {Gill}, Astronomy and Computing \textbf{10}, 121 (2015), arXiv:
  \eprint{1407.7676}.

\bibitem{Davari2016}
R.~{Davari}, L.~C. {Ho}, and C.~Y. {Peng}, \apj \textbf{824}, 112 (2016),
  arXiv: \eprint{1604.08331}.

\bibitem{Davari2017}
R.~H. {Davari}, L.~C. {Ho}, B.~{Mobasher}, and G.~{Canalizo}, \apj
  \textbf{836}, 75 (2017), arXiv: \eprint{1606.07571}.

\bibitem{Euclid2023XXVI}
{Euclid Collaboration}, H.~{Bretonni{\`e}re}, U.~{Kuchner},
  M.~{Huertas-Company}, E.~{Merlin}, M.~{Castellano}, D.~{Tuccillo},
  F.~{Buitrago}, C.~J. {Conselice}, A.~{Boucaud}, B.~{H{\"a}u{\ss}ler},
  M.~{K{\"u}mmel}, W.~G. {Hartley}, A.~{Alvarez Ayllon}, E.~{Bertin},
  F.~{Ferrari}, L.~{Ferreira}, R.~{Gavazzi}, D.~{Hern{\'a}ndez-Lang},
  G.~{Lucatelli}, A.~S.~G. {Robotham}, M.~{Schefer}, L.~{Wang}, R.~{Cabanac},
  H.~{Dom{\'\i}nguez S{\'a}nchez}, P.~A. {Duc}, S.~{Fotopoulou}, S.~{Kruk},
  A.~{La Marca}, B.~{Margalef-Bentabol}, F.~R. {Marleau}, C.~{Tortora},
  N.~{Aghanim}, A.~{Amara}, N.~{Auricchio}, R.~{Azzollini}, M.~{Baldi},
  R.~{Bender}, C.~{Bodendorf}, E.~{Branchini}, M.~{Brescia}, J.~{Brinchmann},
  S.~{Camera}, V.~{Capobianco}, C.~{Carbone}, J.~{Carretero}, F.~J.
  {Castander}, S.~{Cavuoti}, A.~{Cimatti}, R.~{Cledassou}, G.~{Congedo},
  L.~{Conversi}, Y.~{Copin}, L.~{Corcione}, F.~{Courbin}, M.~{Cropper}, A.~{Da
  Silva}, H.~{Degaudenzi}, J.~{Dinis}, F.~{Dubath}, C.~A.~J. {Duncan},
  X.~{Dupac}, S.~{Dusini}, S.~{Farrens}, S.~{Ferriol}, M.~{Frailis},
  E.~{Franceschi}, M.~{Fumana}, S.~{Galeotta}, B.~{Garilli}, B.~{Gillis},
  C.~{Giocoli}, A.~{Grazian}, F.~{Grupp}, S.~V.~H. {Haugan}, H.~{Hoekstra},
  W.~{Holmes}, F.~{Hormuth}, A.~{Hornstrup}, P.~{Hudelot}, K.~{Jahnke},
  S.~{Kermiche}, A.~{Kiessling}, R.~{Kohley}, M.~{Kunz}, H.~{Kurki-Suonio},
  S.~{Ligori}, P.~B. {Lilje}, I.~{Lloro}, O.~{Mansutti}, O.~{Marggraf},
  K.~{Markovic}, F.~{Marulli}, R.~{Massey}, H.~J. {McCracken}, E.~{Medinaceli},
  M.~{Melchior}, M.~{Meneghetti}, G.~{Meylan}, M.~{Moresco}, L.~{Moscardini},
  E.~{Munari}, S.~M. {Niemi}, C.~{Padilla}, S.~{Paltani}, F.~{Pasian},
  K.~{Pedersen}, W.~{Percival}, V.~{Pettorino}, G.~{Polenta}, M.~{Poncet},
  L.~{Pozzetti}, F.~{Raison}, R.~{Rebolo}, A.~{Renzi}, J.~{Rhodes},
  G.~{Riccio}, E.~{Romelli}, C.~{Rosset}, E.~{Rossetti}, R.~{Saglia},
  D.~{Sapone}, B.~{Sartoris}, P.~{Schneider}, A.~{Secroun}, G.~{Seidel},
  C.~{Sirignano}, G.~{Sirri}, J.~{Skottfelt}, J.~L. {Starck},
  P.~{Tallada-Cresp{\'\i}}, A.~N. {Taylor}, I.~{Tereno}, R.~{Toledo-Moreo},
  I.~{Tutusaus}, E.~A. {Valentijn}, L.~{Valenziano}, T.~{Vassallo}, Y.~{Wang},
  J.~{Weller}, G.~{Zamorani}, J.~{Zoubian}, S.~{Andreon}, S.~{Bardelli},
  C.~{Colodro-Conde}, D.~{Di Ferdinando}, J.~{Graci{\'a}-Carpio},
  V.~{Lindholm}, N.~{Mauri}, S.~{Mei}, V.~{Scottez}, E.~{Zucca},
  C.~{Baccigalupi}, M.~{Ballardini}, F.~{Bernardeau}, A.~{Biviano},
  S.~{Borgani}, A.~S. {Borlaff}, C.~{Burigana}, A.~{Cappi}, C.~S. {Carvalho},
  S.~{Casas}, G.~{Castignani}, A.~R. {Cooray}, J.~{Coupon}, H.~M. {Courtois},
  S.~{Davini}, G.~{De Lucia}, G.~{Desprez}, J.~A. {Escartin}, S.~{Escoffier},
  M.~{Fabricius}, M.~{Farina}, A.~{Fontana}, K.~{Ganga}, J.~{Garcia-Bellido},
  K.~{George}, G.~{Gozaliasl}, H.~{Hildebrandt}, I.~{Hook}, O.~{Ilbert},
  S.~{Ili{\'c}}, B.~{Joachimi}, V.~{Kansal}, E.~{Keihanen}, C.~C.
  {Kirkpatrick}, A.~{Loureiro}, J.~{Macias-Perez}, M.~{Magliocchetti},
  R.~{Maoli}, S.~{Marcin}, M.~{Martinelli}, N.~{Martinet}, M.~{Maturi},
  P.~{Monaco}, G.~{Morgante}, S.~{Nadathur}, A.~A. {Nucita}, L.~{Patrizii},
  V.~{Popa}, C.~{Porciani}, D.~{Potter}, A.~{Pourtsidou}, M.~{P{\"o}ntinen},
  P.~{Reimberg}, A.~G. {S{\'a}nchez}, Z.~{Sakr}, M.~{Schirmer}, E.~{Sefusatti},
  M.~{Sereno}, J.~{Stadel}, R.~{Teyssier}, J.~{Valiviita}, S.~E. {van Mierlo},
  A.~{Veropalumbo}, M.~{Viel}, J.~R. {Weaver}, and D.~{Scott}, \aap
  \textbf{671}, A102 (2023), arXiv: \eprint{2209.12907}.

\bibitem{Lzy2025}
X.~{Li}, Z.-Y. {Li}, Y.~A. {Li}, M.-Y. {Zhuang}, and X.~{Liao}, \aap
  \textbf{702}, A89 (2025), arXiv: \eprint{2508.06932}.

\bibitem{Giavalisco1996}
M.~{Giavalisco}, M.~{Livio}, R.~C. {Bohlin}, F.~D. {Macchetto}, and T.~P.
  {Stecher}, \aj \textbf{112}, 369 (1996).

\bibitem{Conselice2003}
C.~J. {Conselice}, \apjs \textbf{147}, 1 (2003), arXiv:
  \eprint{astro-ph/0303065}.

\bibitem{Yu2018}
S.-Y. {Yu}, L.~C. {Ho}, A.~J. {Barth}, and Z.-Y. {Li}, \apj \textbf{862}, 13
  (2018), arXiv: \eprint{1806.06591}.

\bibitem{Liang2024}
X.~{Liang}, S.-Y. {Yu}, T.~{Fang}, and L.~C. {Ho}, \aap \textbf{688}, A158
  (2024), arXiv: \eprint{2311.04019}.

\bibitem{Chabrier2003}
G.~{Chabrier}, \pasp \textbf{115}, 763 (2003), arXiv:
  \eprint{astro-ph/0304382}.

\bibitem{Schreiber2017}
C.~{Schreiber}, D.~{Elbaz}, M.~{Pannella}, E.~{Merlin}, M.~{Castellano},
  A.~{Fontana}, N.~{Bourne}, K.~{Boutsia}, F.~{Cullen}, J.~{Dunlop}, H.~C.
  {Ferguson}, M.~J. {Micha{\l}owski}, K.~{Okumura}, P.~{Santini}, X.~W. {Shu},
  T.~{Wang}, and C.~{White}, \aap \textbf{602}, A96 (2017), arXiv:
  \eprint{1606.05354}.

\bibitem{Grogin2011}
N.~A. {Grogin}, D.~D. {Kocevski}, S.~M. {Faber}, H.~C. {Ferguson}, A.~M.
  {Koekemoer}, A.~G. {Riess}, V.~{Acquaviva}, D.~M. {Alexander}, O.~{Almaini},
  M.~L.~N. {Ashby}, M.~{Barden}, E.~F. {Bell}, F.~{Bournaud}, T.~M. {Brown},
  K.~I. {Caputi}, S.~{Casertano}, P.~{Cassata}, M.~{Castellano}, P.~{Challis},
  R.-R. {Chary}, E.~{Cheung}, M.~{Cirasuolo}, C.~J. {Conselice}, A.~{Roshan
  Cooray}, D.~J. {Croton}, E.~{Daddi}, T.~{Dahlen}, R.~{Dav{\'e}}, D.~F. {de
  Mello}, A.~{Dekel}, M.~{Dickinson}, T.~{Dolch}, J.~L. {Donley}, J.~S.
  {Dunlop}, A.~A. {Dutton}, D.~{Elbaz}, G.~G. {Fazio}, A.~V. {Filippenko},
  S.~L. {Finkelstein}, A.~{Fontana}, J.~P. {Gardner}, P.~M. {Garnavich},
  E.~{Gawiser}, M.~{Giavalisco}, A.~{Grazian}, Y.~{Guo}, N.~P. {Hathi},
  B.~{H{\"a}ussler}, P.~F. {Hopkins}, J.-S. {Huang}, K.-H. {Huang}, S.~W.
  {Jha}, J.~S. {Kartaltepe}, R.~P. {Kirshner}, D.~C. {Koo}, K.~{Lai}, K.-S.
  {Lee}, W.~{Li}, J.~M. {Lotz}, R.~A. {Lucas}, P.~{Madau}, P.~J. {McCarthy},
  E.~J. {McGrath}, D.~H. {McIntosh}, R.~J. {McLure}, B.~{Mobasher}, L.~A.
  {Moustakas}, M.~{Mozena}, K.~{Nandra}, J.~A. {Newman}, S.-M. {Niemi}, K.~G.
  {Noeske}, C.~J. {Papovich}, L.~{Pentericci}, A.~{Pope}, J.~R. {Primack},
  A.~{Rajan}, S.~{Ravindranath}, N.~A. {Reddy}, A.~{Renzini}, H.-W. {Rix},
  A.~R. {Robaina}, S.~A. {Rodney}, D.~J. {Rosario}, P.~{Rosati},
  S.~{Salimbeni}, C.~{Scarlata}, B.~{Siana}, L.~{Simard}, J.~{Smidt}, R.~S.
  {Somerville}, H.~{Spinrad}, A.~N. {Straughn}, L.-G. {Strolger}, O.~{Telford},
  H.~I. {Teplitz}, J.~R. {Trump}, A.~{van der Wel}, C.~{Villforth}, R.~H.
  {Wechsler}, B.~J. {Weiner}, T.~{Wiklind}, V.~{Wild}, G.~{Wilson}, S.~{Wuyts},
  H.-J. {Yan}, and M.~S. {Yun}, \apjs \textbf{197}, 35 (2011), arXiv:
  \eprint{1105.3753}.

\bibitem{Koekemoer2011}
A.~M. {Koekemoer}, S.~M. {Faber}, H.~C. {Ferguson}, N.~A. {Grogin}, D.~D.
  {Kocevski}, D.~C. {Koo}, K.~{Lai}, J.~M. {Lotz}, R.~A. {Lucas}, E.~J.
  {McGrath}, S.~{Ogaz}, A.~{Rajan}, A.~G. {Riess}, S.~A. {Rodney},
  L.~{Strolger}, S.~{Casertano}, M.~{Castellano}, T.~{Dahlen}, M.~{Dickinson},
  T.~{Dolch}, A.~{Fontana}, M.~{Giavalisco}, A.~{Grazian}, Y.~{Guo}, N.~P.
  {Hathi}, K.-H. {Huang}, A.~{van der Wel}, H.-J. {Yan}, V.~{Acquaviva}, D.~M.
  {Alexander}, O.~{Almaini}, M.~L.~N. {Ashby}, M.~{Barden}, E.~F. {Bell},
  F.~{Bournaud}, T.~M. {Brown}, K.~I. {Caputi}, P.~{Cassata}, P.~J. {Challis},
  R.-R. {Chary}, E.~{Cheung}, M.~{Cirasuolo}, C.~J. {Conselice}, A.~{Roshan
  Cooray}, D.~J. {Croton}, E.~{Daddi}, R.~{Dav{\'e}}, D.~F. {de Mello}, L.~{de
  Ravel}, A.~{Dekel}, J.~L. {Donley}, J.~S. {Dunlop}, A.~A. {Dutton},
  D.~{Elbaz}, G.~G. {Fazio}, A.~V. {Filippenko}, S.~L. {Finkelstein},
  C.~{Frazer}, J.~P. {Gardner}, P.~M. {Garnavich}, E.~{Gawiser},
  R.~{Gruetzbauch}, W.~G. {Hartley}, B.~{H{\"a}ussler}, J.~{Herrington}, P.~F.
  {Hopkins}, J.-S. {Huang}, S.~W. {Jha}, A.~{Johnson}, J.~S. {Kartaltepe},
  A.~A. {Khostovan}, R.~P. {Kirshner}, C.~{Lani}, K.-S. {Lee}, W.~{Li},
  P.~{Madau}, P.~J. {McCarthy}, D.~H. {McIntosh}, R.~J. {McLure},
  C.~{McPartland}, B.~{Mobasher}, H.~{Moreira}, A.~{Mortlock}, L.~A.
  {Moustakas}, M.~{Mozena}, K.~{Nandra}, J.~A. {Newman}, J.~L. {Nielsen},
  S.~{Niemi}, K.~G. {Noeske}, C.~J. {Papovich}, L.~{Pentericci}, A.~{Pope},
  J.~R. {Primack}, S.~{Ravindranath}, N.~A. {Reddy}, A.~{Renzini}, H.-W. {Rix},
  A.~R. {Robaina}, D.~J. {Rosario}, P.~{Rosati}, S.~{Salimbeni}, C.~{Scarlata},
  B.~{Siana}, L.~{Simard}, J.~{Smidt}, D.~{Snyder}, R.~S. {Somerville},
  H.~{Spinrad}, A.~N. {Straughn}, O.~{Telford}, H.~I. {Teplitz}, J.~R. {Trump},
  C.~{Vargas}, C.~{Villforth}, C.~R. {Wagner}, P.~{Wandro}, R.~H. {Wechsler},
  B.~J. {Weiner}, T.~{Wiklind}, V.~{Wild}, G.~{Wilson}, S.~{Wuyts}, and M.~S.
  {Yun}, \apjs \textbf{197}, 36 (2011), arXiv: \eprint{1105.3754}.

\bibitem{Guo2013}
Y.~{Guo}, H.~C. {Ferguson}, M.~{Giavalisco}, G.~{Barro}, S.~P. {Willner},
  M.~L.~N. {Ashby}, T.~{Dahlen}, J.~L. {Donley}, S.~M. {Faber}, A.~{Fontana},
  A.~{Galametz}, A.~{Grazian}, K.-H. {Huang}, D.~D. {Kocevski}, A.~M.
  {Koekemoer}, D.~C. {Koo}, E.~J. {McGrath}, M.~{Peth}, M.~{Salvato},
  S.~{Wuyts}, M.~{Castellano}, A.~R. {Cooray}, M.~E. {Dickinson}, J.~S.
  {Dunlop}, G.~G. {Fazio}, J.~P. {Gardner}, E.~{Gawiser}, N.~A. {Grogin}, N.~P.
  {Hathi}, L.-T. {Hsu}, K.-S. {Lee}, R.~A. {Lucas}, B.~{Mobasher}, K.~{Nandra},
  J.~A. {Newman}, and A.~{van der Wel}, \apjs \textbf{207}, 24 (2013), arXiv:
  \eprint{1308.4405}.

\bibitem{Galametz2013}
A.~{Galametz}, A.~{Grazian}, A.~{Fontana}, H.~C. {Ferguson}, M.~L.~N. {Ashby},
  G.~{Barro}, M.~{Castellano}, T.~{Dahlen}, J.~L. {Donley}, S.~M. {Faber},
  N.~{Grogin}, Y.~{Guo}, K.-H. {Huang}, D.~D. {Kocevski}, A.~M. {Koekemoer},
  K.-S. {Lee}, E.~J. {McGrath}, M.~{Peth}, S.~P. {Willner}, O.~{Almaini},
  M.~{Cooper}, A.~{Cooray}, C.~J. {Conselice}, M.~{Dickinson}, J.~S. {Dunlop},
  G.~G. {Fazio}, S.~{Foucaud}, J.~P. {Gardner}, M.~{Giavalisco}, N.~P. {Hathi},
  W.~G. {Hartley}, D.~C. {Koo}, K.~{Lai}, D.~F. {de Mello}, R.~J. {McLure},
  R.~A. {Lucas}, D.~{Paris}, L.~{Pentericci}, P.~{Santini}, C.~{Simpson},
  V.~{Sommariva}, T.~{Targett}, B.~J. {Weiner}, S.~{Wuyts}, and {CANDELS Team},
  \apjs \textbf{206}, 10 (2013), arXiv: \eprint{1305.1823}.

\bibitem{Nayyeri2017}
H.~{Nayyeri}, S.~{Hemmati}, B.~{Mobasher}, H.~C. {Ferguson}, A.~{Cooray},
  G.~{Barro}, S.~M. {Faber}, M.~{Dickinson}, A.~M. {Koekemoer}, M.~{Peth},
  M.~{Salvato}, M.~L.~N. {Ashby}, B.~{Darvish}, J.~{Donley}, M.~{Durbin},
  S.~{Finkelstein}, A.~{Fontana}, N.~A. {Grogin}, R.~{Gruetzbauch}, K.~{Huang},
  A.~A. {Khostovan}, D.~{Kocevski}, D.~{Kodra}, B.~{Lee}, J.~{Newman},
  C.~{Pacifici}, J.~{Pforr}, M.~{Stefanon}, T.~{Wiklind}, S.~P. {Willner},
  S.~{Wuyts}, M.~{Castellano}, C.~{Conselice}, T.~{Dolch}, J.~S. {Dunlop},
  A.~{Galametz}, N.~P. {Hathi}, R.~A. {Lucas}, and H.~{Yan}, \apjs
  \textbf{228}, 7 (2017), arXiv: \eprint{1612.07364}.

\bibitem{Lang2014}
P.~{Lang}, S.~{Wuyts}, R.~S. {Somerville}, N.~M. {F{\"o}rster Schreiber},
  R.~{Genzel}, E.~F. {Bell}, G.~{Brammer}, A.~{Dekel}, S.~M. {Faber}, H.~C.
  {Ferguson}, N.~A. {Grogin}, D.~D. {Kocevski}, A.~M. {Koekemoer}, D.~{Lutz},
  E.~J. {McGrath}, I.~{Momcheva}, E.~J. {Nelson}, J.~R. {Primack}, D.~J.
  {Rosario}, R.~E. {Skelton}, L.~J. {Tacconi}, P.~G. {van Dokkum}, and K.~E.
  {Whitaker}, \apj \textbf{788}, 11 (2014), arXiv: \eprint{1402.0866}.

\bibitem{Han2025}
J.~{Han}, M.~{Li}, W.~{Jiang}, Z.~{Chen}, H.~{Wang}, C.~{Wei}, F.~{He},
  J.~{He}, J.~{Zhang}, Y.~{Liu}, W.~{Cui}, Y.~{Gu}, Q.~{Guo}, Y.~{Jing},
  X.~{Kang}, G.~{Li}, X.~{Luo}, Y.~{Luo}, W.~{Pei}, Y.~{Qiu}, Z.~{Tan},
  L.~{Xie}, X.~{Yang}, H.~{Yu}, Y.~{Yu}, and J.~{Zhou}, Science China Physics,
  Mechanics, and Astronomy \textbf{68}, 109511 (2025), arXiv:
  \eprint{2503.21368}.

\bibitem{Zhang2019RAA}
Y.-C. {Zhang} and X.-H. {Yang}, Research in Astronomy and Astrophysics
  \textbf{19}, 006 (2019), arXiv: \eprint{1707.04979}.

\bibitem{Hiemer2014}
A.~{Hiemer}, M.~{Barden}, L.~S. {Kelvin}, B.~{H{\"a}u{\ss}ler}, and
  S.~{Schindler}, \mnras \textbf{444}, 3089 (2014).

\bibitem{Euclid2023Merlin}
{Euclid Collaboration}, E.~{Merlin}, M.~{Castellano}, H.~{Bretonni{\`e}re},
  M.~{Huertas-Company}, U.~{Kuchner}, D.~{Tuccillo}, F.~{Buitrago}, J.~R.
  {Peterson}, C.~J. {Conselice}, F.~{Caro}, P.~{Dimauro}, L.~{Nemani},
  A.~{Fontana}, M.~{K{\"u}mmel}, B.~{H{\"a}u{\ss}ler}, W.~G. {Hartley},
  A.~{Alvarez Ayllon}, E.~{Bertin}, P.~{Dubath}, F.~{Ferrari}, L.~{Ferreira},
  R.~{Gavazzi}, D.~{Hern{\'a}ndez-Lang}, G.~{Lucatelli}, A.~S.~G. {Robotham},
  M.~{Schefer}, C.~{Tortora}, N.~{Aghanim}, A.~{Amara}, L.~{Amendola},
  N.~{Auricchio}, M.~{Baldi}, R.~{Bender}, C.~{Bodendorf}, E.~{Branchini},
  M.~{Brescia}, S.~{Camera}, V.~{Capobianco}, C.~{Carbone}, J.~{Carretero},
  F.~J. {Castander}, S.~{Cavuoti}, A.~{Cimatti}, R.~{Cledassou}, G.~{Congedo},
  L.~{Conversi}, Y.~{Copin}, L.~{Corcione}, F.~{Courbin}, M.~{Cropper}, A.~{Da
  Silva}, H.~{Degaudenzi}, J.~{Dinis}, M.~{Douspis}, F.~{Dubath}, C.~A.~J.
  {Duncan}, X.~{Dupac}, S.~{Dusini}, S.~{Farrens}, S.~{Ferriol}, M.~{Frailis},
  E.~{Franceschi}, P.~{Franzetti}, S.~{Galeotta}, B.~{Garilli}, B.~{Gillis},
  C.~{Giocoli}, A.~{Grazian}, F.~{Grupp}, S.~V.~H. {Haugan}, H.~{Hoekstra},
  W.~{Holmes}, F.~{Hormuth}, A.~{Hornstrup}, P.~{Hudelot}, K.~{Jahnke},
  S.~{Kermiche}, A.~{Kiessling}, T.~{Kitching}, R.~{Kohley}, M.~{Kunz},
  H.~{Kurki-Suonio}, S.~{Ligori}, P.~B. {Lilje}, I.~{Lloro}, O.~{Mansutti},
  O.~{Marggraf}, K.~{Markovic}, F.~{Marulli}, R.~{Massey}, H.~J. {McCracken},
  E.~{Medinaceli}, M.~{Melchior}, M.~{Meneghetti}, G.~{Meylan}, M.~{Moresco},
  L.~{Moscardini}, E.~{Munari}, S.~M. {Niemi}, C.~{Padilla}, S.~{Paltani},
  F.~{Pasian}, K.~{Pedersen}, W.~J. {Percival}, G.~{Polenta}, M.~{Poncet},
  L.~{Popa}, L.~{Pozzetti}, F.~{Raison}, R.~{Rebolo}, A.~{Renzi}, J.~{Rhodes},
  G.~{Riccio}, E.~{Romelli}, E.~{Rossetti}, R.~{Saglia}, D.~{Sapone},
  B.~{Sartoris}, P.~{Schneider}, A.~{Secroun}, G.~{Seidel}, C.~{Sirignano},
  G.~{Sirri}, J.~{Skottfelt}, J.-L. {Starck}, P.~{Tallada-Cresp{\'\i}}, A.~N.
  {Taylor}, I.~{Tereno}, R.~{Toledo-Moreo}, I.~{Tutusaus}, L.~{Valenziano},
  T.~{Vassallo}, Y.~{Wang}, J.~{Weller}, A.~{Zacchei}, G.~{Zamorani},
  J.~{Zoubian}, S.~{Andreon}, S.~{Bardelli}, A.~{Boucaud}, C.~{Colodro-Conde},
  D.~{Di Ferdinando}, J.~{Graci{\'a}-Carpio}, V.~{Lindholm}, N.~{Mauri},
  S.~{Mei}, C.~{Neissner}, V.~{Scottez}, A.~{Tramacere}, E.~{Zucca},
  C.~{Baccigalupi}, A.~{Balaguera-Antol{\'\i}nez}, M.~{Ballardini},
  F.~{Bernardeau}, A.~{Biviano}, S.~{Borgani}, A.~S. {Borlaff}, C.~{Burigana},
  R.~{Cabanac}, A.~{Cappi}, C.~S. {Carvalho}, S.~{Casas}, G.~{Castignani},
  A.~R. {Cooray}, J.~{Coupon}, H.~M. {Courtois}, O.~{Cucciati}, S.~{Davini},
  G.~{De Lucia}, G.~{Desprez}, J.~A. {Escartin}, S.~{Escoffier}, M.~{Farina},
  K.~{Ganga}, J.~{Garcia-Bellido}, K.~{George}, G.~{Gozaliasl},
  H.~{Hildebrandt}, I.~{Hook}, O.~{Ilbert}, S.~{Ili{\'c}}, B.~{Joachimi},
  V.~{Kansal}, E.~{Keihanen}, C.~C. {Kirkpatrick}, A.~{Loureiro},
  J.~{Macias-Perez}, M.~{Magliocchetti}, G.~{Mainetti}, R.~{Maoli},
  S.~{Marcin}, M.~{Martinelli}, N.~{Martinet}, S.~{Matthew}, M.~{Maturi}, R.~B.
  {Metcalf}, P.~{Monaco}, G.~{Morgante}, and S.~{Nadathur}, \aap \textbf{671},
  A101 (2023), arXiv: \eprint{2209.12906}.

\bibitem{larry_bradley_2025}
L.~Bradley, B.~Sip{\H o}cz, T.~Robitaille, E.~Tollerud, Z.~Vin{\'{\i}}cius,
  C.~Deil, K.~Barbary, T.~J. Wilson, I.~Busko, A.~Donath, H.~M. G{\"u}nther,
  M.~Cara, P.~L. Lim, S.~Me{\ss}linger, Z.~Burnett, S.~Conseil, M.~Droettboom,
  A.~Bostroem, E.~M. Bray, L.~A. Bratholm, W.~Jamieson, A.~Ginsburg,
  G.~Barentsen, M.~Craig, S.~Pascual, S.~Rathi, M.~Perrin, and B.~M. Morris,
  astropy/photutils: 2.2.0 (2025),
  \urlprefix\url{https://doi.org/10.5281/zenodo.14889440}.

\bibitem{Sersic1968}
J.~L. {Sersic}, \emph{{Atlas de Galaxias Australes}} (1968).

\bibitem{Ferreira2022b}
L.~{Ferreira}, C.~J. {Conselice}, E.~{Sazonova}, F.~{Ferrari}, J.~{Caruana},
  C.-B. {Tohill}, G.~{Lucatelli}, N.~{Adams}, D.~{Irodotou}, M.~A. {Marshall},
  W.~J. {Roper}, C.~C. {Lovell}, A.~{Verma}, D.~{Austin}, J.~{Trussler}, and
  S.~M. {Wilkins}, arXiv e-prints arXiv:2210.01110 (2022), arXiv:
  \eprint{2210.01110}.

\bibitem{Nelson2023}
E.~J. {Nelson}, K.~A. {Suess}, R.~{Bezanson}, S.~H. {Price}, P.~{van Dokkum},
  J.~{Leja}, B.~{Wang}, K.~E. {Whitaker}, I.~{Labb{\'e}}, L.~{Barrufet},
  G.~{Brammer}, D.~J. {Eisenstein}, J.~{Gibson}, A.~I. {Hartley}, B.~D.
  {Johnson}, K.~E. {Heintz}, E.~{Mathews}, T.~B. {Miller}, P.~A. {Oesch},
  L.~{Sandles}, D.~J. {Setton}, J.~S. {Speagle}, S.~{Tacchella}, K.-i.
  {Tadaki}, H.~{{\"U}bler}, and J.~R. {Weaver}, \apjl \textbf{948}, L18 (2023),
  arXiv: \eprint{2208.01630}.

\bibitem{Robertson2023}
B.~E. {Robertson}, S.~{Tacchella}, B.~D. {Johnson}, R.~{Hausen}, A.~B. {Alabi},
  K.~{Boyett}, A.~J. {Bunker}, S.~{Carniani}, E.~{Egami}, D.~J. {Eisenstein},
  K.~N. {Hainline}, J.~M. {Helton}, Z.~{Ji}, N.~{Kumari}, J.~{Lyu},
  R.~{Maiolino}, E.~J. {Nelson}, M.~J. {Rieke}, I.~{Shivaei}, F.~{Sun},
  H.~{{\"U}bler}, C.~C. {Williams}, C.~N.~A. {Willmer}, and J.~{Witstok}, \apjl
  \textbf{942}, L42 (2023), arXiv: \eprint{2208.11456}.

\bibitem{Jacobs2023}
C.~{Jacobs}, K.~{Glazebrook}, A.~{Calabr{\`o}}, T.~{Treu}, T.~{Nannayakkara},
  T.~{Jones}, E.~{Merlin}, R.~{Abraham}, A.~R.~H. {Stevens}, B.~{Vulcani},
  L.~{Yang}, A.~{Bonchi}, K.~{Boyett}, M.~{Brada{\v{c}}}, M.~{Castellano},
  A.~{Fontana}, D.~{Marchesini}, M.~{Malkan}, C.~{Mason}, T.~{Morishita},
  D.~{Paris}, P.~{Santini}, M.~{Trenti}, and X.~{Wang}, \apjl \textbf{948}, L13
  (2023), arXiv: \eprint{2208.06516}.

\bibitem{Cheng2022b}
C.~{Cheng}, H.~{Yan}, J.-S. {Huang}, C.~N.~A. {Willmer}, Z.~{Ma}, and
  G.~{Orellana-Gonz{\'a}lez}, \apjl \textbf{936}, L19 (2022), arXiv:
  \eprint{2207.08234}.

\bibitem{Cheng2023}
C.~{Cheng}, J.-S. {Huang}, I.~{Smail}, H.~{Yan}, S.~H. {Cohen}, R.~A. {Jansen},
  R.~A. {Windhorst}, Z.~{Ma}, A.~{Koekemoer}, C.~N.~A. {Willmer}, S.~P.
  {Willner}, J.~M. {Diego}, B.~{Frye}, C.~J. {Conselice}, L.~{Ferreira},
  A.~{Petric}, M.~{Yun}, H.~B. {Gim}, M.~d.~C. {Polletta}, K.~J. {Duncan},
  B.~W. {Holwerda}, H.~J.~A. {R{\"o}ttgering}, R.~{Honor}, N.~P. {Hathi}, P.~S.
  {Kamieneski}, N.~J. {Adams}, D.~{Coe}, T.~{Broadhurst}, J.~{Summers},
  S.~{Tompkins}, S.~P. {Driver}, N.~A. {Grogin}, M.~A. {Marshall},
  N.~{Pirzkal}, A.~{Robotham}, and R.~E. {Ryan}, \apjl \textbf{942}, L19
  (2023), arXiv: \eprint{2210.08163}.

\bibitem{LeConte2024}
Z.~A. {Le Conte}, D.~A. {Gadotti}, L.~{Ferreira}, C.~J. {Conselice}, C.~{de
  S{\'a}-Freitas}, T.~{Kim}, J.~{Neumann}, F.~{Fragkoudi}, E.~{Athanassoula},
  and N.~J. {Adams}, \mnras \textbf{530}, 1984 (2024), arXiv:
  \eprint{2309.10038}.

\bibitem{XuYu2024}
D.~{Xu} and S.-Y. {Yu}, \aap \textbf{682}, L17 (2024), arXiv:
  \eprint{2402.04233}.

\bibitem{Yu2025}
S.-Y. {Yu}, D.~{Xu}, B.~S. {Kalita}, S.~{Li}, J.~D. {Silverman}, X.~{Liang},
  and T.~{Fang}, \aap \textbf{693}, L9 (2025), arXiv: \eprint{2412.13064}.

\bibitem{Yu2026}
S.-Y. {Yu}, L.~C. {Ho}, T.~{Tsukui}, J.~D. {Silverman}, M.~{Huertas-Company},
  A.~M. {Koekemoer}, M.~{Franco}, R.~{Massey}, L.~{Yang}, R.~C. {Arango-Toro},
  A.~L. {Faisst}, G.~{Gozaliasl}, K.~{Sheth}, J.~S. {Kartaltepe}, C.~{Xu},
  A.~{Haghjoo}, X.~{Ding}, Z.~{Liu}, and J.~{McCleary}, arXiv e-prints
  arXiv:2601.04988 (2026), arXiv: \eprint{2601.04988}.

\bibitem{Huertas-Company2025}
M.~{Huertas-Company}, M.~{Shuntov}, Y.~{Dong}, M.~{Walmsley}, O.~{Ilbert},
  H.~J. {McCracken}, H.~B. {Akins}, N.~{Allen}, C.~M. {Casey}, L.~{Costantin},
  E.~{Daddi}, A.~{Dekel}, M.~{Franco}, I.~L. {Garland}, T.~{G{\'e}ron},
  G.~{Gozaliasl}, M.~{Hirschmann}, J.~S. {Kartaltepe}, A.~M. {Koekemoer},
  C.~{Lintott}, D.~{Liu}, R.~{Lucas}, K.~{Masters}, F.~{Pacucci},
  L.~{Paquereau}, P.~G. {P'erez-Gonz'alez}, J.~D. {Rhodes}, B.~E. {Robertson},
  B.~{Simmons}, R.~{Smethurst}, S.~{Toft}, and L.~{Yang}, arXiv e-prints
  arXiv:2502.03532 (2025), arXiv: \eprint{2502.03532}.

\bibitem{Stone2023}
C.~J. {Stone}, S.~{Courteau}, J.-C. {Cuillandre}, Y.~{Hezaveh},
  L.~{Perreault-Levasseur}, and N.~{Arora}, \mnras \textbf{525}, 6377 (2023),
  arXiv: \eprint{2308.01957}.

\bibitem{Bertin1996}
E.~{Bertin} and S.~{Arnouts}, \aaps \textbf{117}, 393 (1996).

\bibitem{Barbary2016}
K.~Barbary, Journal of Open Source Software \textbf{1}, 58 (2016),
  \urlprefix\url{https://doi.org/10.21105/joss.00058}.

\bibitem{Peng2002}
C.~Y. {Peng}, L.~C. {Ho}, C.~D. {Impey}, and H.-W. {Rix}, \aj \textbf{124}, 266
  (2002), arXiv: \eprint{astro-ph/0204182}.

\bibitem{Peng2010}
C.~Y. {Peng}, L.~C. {Ho}, C.~D. {Impey}, and H.-W. {Rix}, \aj \textbf{139},
  2097 (2010), arXiv: \eprint{0912.0731}.

\bibitem{Bertin2020}
E.~{Bertin}, M.~{Schefer}, N.~{Apostolakos}, A.~{{\'A}lvarez-Ayll{\'o}n},
  P.~{Dubath}, and M.~{K{\"u}mmel}, in \emph{Astronomical Data Analysis
  Software and Systems XXIX}, (edited by R.~{Pizzo}, E.~R. {Deul}, J.~D. {Mol},
  J.~{de Plaa}, and H.~{Verkouter}), volume 527 of \emph{Astronomical Society
  of the Pacific Conference Series}, 461 (2020).

\bibitem{Bouwens2004}
R.~J. {Bouwens}, G.~D. {Illingworth}, J.~P. {Blakeslee}, T.~J. {Broadhurst},
  and M.~{Franx}, \apjl \textbf{611}, L1 (2004), arXiv:
  \eprint{astro-ph/0406562}.

\bibitem{Daddi2005}
E.~{Daddi}, A.~{Renzini}, N.~{Pirzkal}, A.~{Cimatti}, S.~{Malhotra},
  M.~{Stiavelli}, C.~{Xu}, A.~{Pasquali}, J.~E. {Rhoads}, M.~{Brusa}, S.~{di
  Serego Alighieri}, H.~C. {Ferguson}, A.~M. {Koekemoer}, L.~A. {Moustakas},
  N.~{Panagia}, and R.~A. {Windhorst}, \apj \textbf{626}, 680 (2005), arXiv:
  \eprint{astro-ph/0503102}.

\bibitem{Trujillo2007}
I.~{Trujillo}, C.~J. {Conselice}, K.~{Bundy}, M.~C. {Cooper}, P.~{Eisenhardt},
  and R.~S. {Ellis}, \mnras \textbf{382}, 109 (2007), arXiv:
  \eprint{0709.0621}.

\bibitem{Buitrago2008}
F.~{Buitrago}, I.~{Trujillo}, C.~J. {Conselice}, R.~J. {Bouwens},
  M.~{Dickinson}, and H.~{Yan}, \apjl \textbf{687}, L61 (2008), arXiv:
  \eprint{0807.4141}.

\bibitem{Oesch2010}
P.~A. {Oesch}, R.~J. {Bouwens}, C.~M. {Carollo}, G.~D. {Illingworth},
  M.~{Trenti}, M.~{Stiavelli}, D.~{Magee}, I.~{Labb{\'e}}, and M.~{Franx},
  \apjl \textbf{709}, L21 (2010), arXiv: \eprint{0909.5183}.

\bibitem{Mosleh2012}
M.~{Mosleh}, R.~J. {Williams}, M.~{Franx}, V.~{Gonzalez}, R.~J. {Bouwens},
  P.~{Oesch}, I.~{Labbe}, G.~D. {Illingworth}, and M.~{Trenti}, \apjl
  \textbf{756}, L12 (2012), arXiv: \eprint{1207.6634}.

\bibitem{Whitney2019}
A.~{Whitney}, C.~J. {Conselice}, R.~{Bhatawdekar}, and K.~{Duncan}, \apj
  \textbf{887}, 113 (2019), arXiv: \eprint{1911.02589}.

\bibitem{Allen2025}
N.~{Allen}, P.~A. {Oesch}, S.~{Toft}, J.~{Matharu}, C.~J.~R. {McPartland},
  A.~{Weibel}, G.~{Brammer}, R.~A.~A. {Bowler}, K.~{Ito}, R.~{Gottumukkala},
  F.~{Rizzo}, F.~{Valentino}, R.~G. {Varadaraj}, J.~R. {Weaver}, and K.~E.
  {Whitaker}, \aap \textbf{698}, A30 (2025), arXiv: \eprint{2410.16354}.

\bibitem{Schade1995}
D.~{Schade}, S.~J. {Lilly}, D.~{Crampton}, F.~{Hammer}, O.~{Le Fevre}, and
  L.~{Tresse}, \apjl \textbf{451}, L1 (1995), arXiv: \eprint{astro-ph/9507028}.

\bibitem{Schade1996}
D.~{Schade}, S.~J. {Lilly}, O.~{Le Fevre}, F.~{Hammer}, and D.~{Crampton}, \apj
  \textbf{464}, 79 (1996), arXiv: \eprint{astro-ph/9601047}.

\bibitem{Lilly1998}
S.~{Lilly}, D.~{Schade}, R.~{Ellis}, O.~{Le F{\`e}vre}, J.~{Brinchmann},
  L.~{Tresse}, R.~{Abraham}, F.~{Hammer}, D.~{Crampton}, M.~{Colless},
  K.~{Glazebrook}, G.~{Mallen-Ornelas}, and T.~{Broadhurst}, \apj \textbf{500},
  75 (1998), arXiv: \eprint{astro-ph/9712061}.

\bibitem{Roche1998}
N.~{Roche}, K.~{Ratnatunga}, R.~E. {Griffiths}, M.~{Im}, and A.~{Naim}, \mnras
  \textbf{293}, 157 (1998).

\bibitem{Labbe2003}
I.~{Labb{\'e}}, G.~{Rudnick}, M.~{Franx}, E.~{Daddi}, P.~G. {van Dokkum}, N.~M.
  {F{\"o}rster Schreiber}, K.~{Kuijken}, A.~{Moorwood}, H.-W. {Rix},
  H.~{R{\"o}ttgering}, I.~{Trujillo}, A.~{van der Wel}, P.~{van der Werf}, and
  L.~{van Starkenburg}, \apjl \textbf{591}, L95 (2003), arXiv:
  \eprint{astro-ph/0306062}.

\bibitem{Barden2005}
M.~{Barden}, H.-W. {Rix}, R.~S. {Somerville}, E.~F. {Bell},
  B.~{H{\"a}u{\ss}ler}, C.~Y. {Peng}, A.~{Borch}, S.~V.~W. {Beckwith}, J.~A.~R.
  {Caldwell}, C.~{Heymans}, K.~{Jahnke}, S.~{Jogee}, D.~H. {McIntosh},
  K.~{Meisenheimer}, S.~F. {S{\'a}nchez}, L.~{Wisotzki}, and C.~{Wolf}, \apj
  \textbf{635}, 959 (2005), arXiv: \eprint{astro-ph/0502416}.

\bibitem{Sobral2013}
D.~{Sobral}, I.~{Smail}, P.~N. {Best}, J.~E. {Geach}, Y.~{Matsuda}, J.~P.
  {Stott}, M.~{Cirasuolo}, and J.~{Kurk}, \mnras \textbf{428}, 1128 (2013),
  arXiv: \eprint{1202.3436}.

\bibitem{Whitney2020}
A.~{Whitney}, C.~J. {Conselice}, K.~{Duncan}, and L.~R. {Spitler}, \apj
  \textbf{903}, 14 (2020), arXiv: \eprint{2009.07295}.

\bibitem{Scoville2007}
N.~{Scoville}, R.~G. {Abraham}, H.~{Aussel}, J.~E. {Barnes}, A.~{Benson}, A.~W.
  {Blain}, D.~{Calzetti}, A.~{Comastri}, P.~{Capak}, C.~{Carilli}, J.~E.
  {Carlstrom}, C.~M. {Carollo}, J.~{Colbert}, E.~{Daddi}, R.~S. {Ellis},
  M.~{Elvis}, S.~P. {Ewald}, M.~{Fall}, A.~{Franceschini}, M.~{Giavalisco},
  W.~{Green}, R.~E. {Griffiths}, L.~{Guzzo}, G.~{Hasinger}, C.~{Impey}, J.-P.
  {Kneib}, J.~{Koda}, A.~{Koekemoer}, O.~{Lefevre}, S.~{Lilly}, C.~T. {Liu},
  H.~J. {McCracken}, R.~{Massey}, Y.~{Mellier}, S.~{Miyazaki}, B.~{Mobasher},
  J.~{Mould}, C.~{Norman}, A.~{Refregier}, A.~{Renzini}, J.~{Rhodes},
  M.~{Rich}, D.~B. {Sanders}, D.~{Schiminovich}, E.~{Schinnerer},
  M.~{Scodeggio}, K.~{Sheth}, P.~L. {Shopbell}, Y.~{Taniguchi}, N.~D. {Tyson},
  C.~M. {Urry}, L.~{Van Waerbeke}, P.~{Vettolani}, S.~D.~M. {White}, and
  L.~{Yan}, \apjs \textbf{172}, 38 (2007), arXiv: \eprint{astro-ph/0612306}.

\bibitem{Koekemoer2007}
A.~M. {Koekemoer}, H.~{Aussel}, D.~{Calzetti}, P.~{Capak}, M.~{Giavalisco},
  J.-P. {Kneib}, A.~{Leauthaud}, O.~{Le F{\`e}vre}, H.~J. {McCracken},
  R.~{Massey}, B.~{Mobasher}, J.~{Rhodes}, N.~{Scoville}, and P.~L. {Shopbell},
  \apjs \textbf{172}, 196 (2007), arXiv: \eprint{astro-ph/0703095}.

\bibitem{Weaver2022}
J.~R. {Weaver}, O.~B. {Kauffmann}, O.~{Ilbert}, H.~J. {McCracken}, A.~{Moneti},
  S.~{Toft}, G.~{Brammer}, M.~{Shuntov}, I.~{Davidzon}, B.~C. {Hsieh},
  C.~{Laigle}, A.~{Anastasiou}, C.~K. {Jespersen}, J.~{Vinther}, P.~{Capak},
  C.~M. {Casey}, C.~J.~R. {McPartland}, B.~{Milvang-Jensen}, B.~{Mobasher},
  D.~B. {Sanders}, L.~{Zalesky}, S.~{Arnouts}, H.~{Aussel}, J.~S. {Dunlop},
  A.~{Faisst}, M.~{Franx}, L.~J. {Furtak}, J.~P.~U. {Fynbo}, K.~M.~L. {Gould},
  T.~R. {Greve}, S.~{Gwyn}, J.~S. {Kartaltepe}, D.~{Kashino}, A.~M.
  {Koekemoer}, V.~{Kokorev}, O.~{Le F{\`e}vre}, S.~{Lilly}, D.~{Masters},
  G.~{Magdis}, V.~{Mehta}, Y.~{Peng}, D.~A. {Riechers}, M.~{Salvato},
  M.~{Sawicki}, C.~{Scarlata}, N.~{Scoville}, R.~{Shirley}, J.~D. {Silverman},
  A.~{Sneppen}, V.~{Smol\v{c}i'{c}}, C.~{Steinhardt}, D.~{Stern}, M.~{Tanaka},
  Y.~{Taniguchi}, H.~I. {Teplitz}, M.~{Vaccari}, W.-H. {Wang}, and
  G.~{Zamorani}, \apjs \textbf{258}, 11 (2022), arXiv: \eprint{2110.13923}.

\bibitem{perez2023}
P.~G. {P{\'e}rez-Gonz{\'a}lez}, G.~{Barro}, M.~{Annunziatella}, L.~{Costantin},
  {\'A}.~{Garc{\'\i}a-Argum{\'a}nez}, E.~J. {McGrath}, R.~M. {M{\'e}rida},
  J.~A. {Zavala}, P.~{Arrabal Haro}, M.~B. {Bagley}, B.~E. {Backhaus},
  P.~{Behroozi}, E.~F. {Bell}, L.~{Bisigello}, V.~{Buat}, A.~{Calabr{\`o}},
  C.~M. {Casey}, N.~J. {Cleri}, R.~T. {Coogan}, M.~C. {Cooper}, A.~R. {Cooray},
  A.~{Dekel}, M.~{Dickinson}, D.~{Elbaz}, H.~C. {Ferguson}, S.~L.
  {Finkelstein}, A.~{Fontana}, M.~{Franco}, J.~P. {Gardner}, M.~{Giavalisco},
  C.~{G{\'o}mez-Guijarro}, A.~{Grazian}, N.~A. {Grogin}, Y.~{Guo},
  M.~{Huertas-Company}, S.~{Jogee}, J.~S. {Kartaltepe}, L.~J. {Kewley},
  A.~{Kirkpatrick}, D.~D. {Kocevski}, A.~M. {Koekemoer}, A.~S. {Long}, J.~M.
  {Lotz}, R.~A. {Lucas}, C.~{Papovich}, N.~{Pirzkal}, S.~{Ravindranath}, R.~S.
  {Somerville}, S.~{Tacchella}, J.~R. {Trump}, W.~{Wang}, S.~M. {Wilkins},
  S.~{Wuyts}, G.~{Yang}, and L.~Y.~A. {Yung}, \apjl \textbf{946}, L16 (2023),
  arXiv: \eprint{2211.00045}.

\bibitem{McKinney2025}
J.~{McKinney}, C.~M. {Casey}, A.~S. {Long}, O.~R. {Cooper}, S.~M. {Manning},
  M.~{Franco}, H.~{Akins}, E.~{Lambrides}, E.~{Gammon}, C.~{Silva},
  F.~{Gentile}, J.~A. {Zavala}, A.~{Amvrosiadis}, I.~{Andika}, M.~{Brinch},
  J.~B. {Champagne}, N.~{Chartab}, N.~E. {Drakos}, A.~L. {Faisst},
  S.~{Fujimoto}, S.~{Gillman}, G.~{Gozaliasl}, T.~R. {Greve}, S.~{Harish},
  C.~C. {Hayward}, M.~{Hirschmann}, O.~{Ilbert}, B.~S. {Kalita}, J.~S.
  {Kartaltepe}, A.~M. {Koekemoer}, V.~{Kokorev}, D.~{Liu}, G.~{Magdis}, H.~J.
  {McCracken}, J.~{Rhodes}, B.~E. {Robertson}, M.~{Talia}, F.~{Valentino}, and
  A.~P. {Vijayan}, \apj \textbf{979}, 229 (2025), arXiv: \eprint{2408.08346}.

\bibitem{Ren2025}
J.~{Ren}, F.~S. {Liu}, N.~{Li}, P.~{Zhao}, Q.~{Cui}, Q.~{Song}, Y.~{Li},
  H.~{Mo}, H.~M. {Yesuf}, W.~{Wang}, F.~{An}, and X.~Z. {Zheng}, \apj
  \textbf{982}, 200 (2025), arXiv: \eprint{2502.15569}.

\bibitem{Guo2015}
Y.~{Guo}, H.~C. {Ferguson}, E.~F. {Bell}, D.~C. {Koo}, C.~J. {Conselice},
  M.~{Giavalisco}, S.~{Kassin}, Y.~{Lu}, R.~{Lucas}, N.~{Mandelker}, D.~H.
  {McIntosh}, J.~R. {Primack}, S.~{Ravindranath}, G.~{Barro}, D.~{Ceverino},
  A.~{Dekel}, S.~M. {Faber}, J.~J. {Fang}, A.~M. {Koekemoer}, K.~{Noeske},
  M.~{Rafelski}, and A.~{Straughn}, \apj \textbf{800}, 39 (2015), arXiv:
  \eprint{1410.7398}.

\bibitem{Bournaud2007}
F.~{Bournaud}, B.~G. {Elmegreen}, and D.~M. {Elmegreen}, \apj \textbf{670}, 237
  (2007), arXiv: \eprint{0708.0306}.

\bibitem{Martorano2025}
M.~{Martorano}, A.~{van der Wel}, M.~{Baes}, E.~F. {Bell}, G.~{Brammer},
  M.~{Franx}, A.~{Gebek}, S.~E. {Meidt}, T.~B. {Miller}, E.~{Nelson},
  A.~{Nersesian}, S.~H. {Price}, P.~{van Dokkum}, K.~E. {Whitaker}, and
  S.~{Wuyts}, \aap \textbf{694}, A76 (2025), arXiv: \eprint{2501.02956}.

\bibitem{Boquien2019}
M.~{Boquien}, D.~{Burgarella}, Y.~{Roehlly}, V.~{Buat}, L.~{Ciesla},
  D.~{Corre}, A.~K. {Inoue}, and H.~{Salas}, \aap \textbf{622}, A103 (2019),
  arXiv: \eprint{1811.03094}.

\end{thebibliography}

%\begin{thebibliography}{99}

%\end{thebibliography}

\end{multicols}
\end{document}